\documentclass{article}
\pdfoutput=1 
\usepackage[paperwidth=8.5in,paperheight=11in,bindingoffset=0in,left=1in,right=1in,bottom=1in,top=1in,footskip=0.37in]{geometry}
\usepackage[sort&compress]{natbib}
\usepackage{graphicx} 
\usepackage[utf8]{inputenc}
\usepackage{setspace}
\usepackage[mathscr]{euscript}
\usepackage{mathrsfs}
\usepackage{amssymb,amsmath,amsthm}
\newtheorem{theorem}{Theorem}
\newtheorem{lemma}{Lemma}
\newtheorem{proposition}{Proposition}
\newtheorem{corollary}{Corollary}
\newtheorem{definition}{Definition}

\newcommand{\ineql}[5]{#1^{#2}{#3_{#5}}\Big(1-#4_{#5}/n\Big)}
\newcommand{\inequ}[5]{#1^{#2}{#3_{#5}}\Big(1+#4_{#5}/n\Big)}

\usepackage{mathtools}
\usepackage{relsize}
\usepackage{amsfonts}
\usepackage{verbatim}
\usepackage{subcaption}
\usepackage{cite}
\usepackage{natbib}
\setcitestyle{numbers}
\setcitestyle{square}
\usepackage{titlesec}
\usepackage{lipsum}
\usepackage[T1]{fontenc}
 \usepackage{relsize}
 \usepackage[hidelinks]{hyperref}
\usepackage[nameinlink]{cleveref}
\renewcommand{\thesection}{\arabic{section}}
\usepackage[all]{nowidow}
\usepackage[activate={true,nocompatibility},final,tracking=true,kerning=true,spacing=true,factor=1100,stretch=10,shrink=10]{microtype}
\microtypecontext{spacing=nonfrench}

\titleformat{\section}
  {\normalfont\centering}{\thesection}{1em}{\uppercase}
  
 \titleformat{\subsection}
  {\normalfont\centering}{\thesubsection}{1em}{\uppercase}
\titleformat{\chapter}[display]
    {\normalfont\centering}{\MakeUppercase{\chaptertitlename} \ \thechapter}{5pt}{\uppercase}
\titlespacing*{\chapter}{0pt}{30pt}{20pt}  
\newcommand{\ditto}[1][.4pt]{~\textquotedbl~}
\newcommand{\thesistitle}{Counting Subnetworks Under Gene Duplication in Genetic Regulatory Networks} 
\newcommand{\nil}[1]{}

\begin{document}
\newpage
\pagenumbering{gobble}
\thispagestyle{empty}
\begin{center}
\textsc{\thesistitle}\\[10pt]
by\\
Ashley Scruse, Jonathan Arnold, Robert Robinson
\end{center}

\begin{abstract} Gene duplication is a fundamental evolutionary mechanism that contributes to biological complexity and diversity \citep{fortna04}. Traditionally, research has focused on the duplication of gene sequences \citep{zhang1914}. However, evidence suggests that the duplication of regulatory elements may also play a significant role in the evolution of genomic functions \citep{sarah, hallin}. In this work the evolution of regulatory relationships belonging to gene-specific-substructures in a GRN are modeled. In the model, a network grows from an initial configuration by repeatedly choosing a random gene to duplicate. The likelihood that the regulatory relationships associated with the selected gene are retained through duplication is determined by a vector of probabilities. That is to say that each gene family has its own probability of retaining regulatory relationships. Occurrences of gene-family-specific substructures are counted under the gene duplication model. In this thesis gene-family-specific substructures are referred to as subnetwork motifs. These subnetwork motifs are motivated by network motifs which are patterns of interconnections that recur more often in a specialized network than in a random network \citep{milo2}. Subnetwork motifs differ from network motifs in the way that subnetwork motifs are instances of gene-family-specific substructures while network motifs are isomorphic substructures. These subnetwork motifs are counted under Full and Partial Duplication, which differ in the way in which regulation relationships are inherited. Full duplication occurs when all regulatory links are inherited at each duplication step, and Partial Duplication occurs when regulation inheritance varies at each duplication step. Note that Full Duplication is just a special case of Partial Duplication. Moments for the number of occurrences of subnetwork motifs are determined in each model. In the end, the results presented offer a method for discovering gene-family-specific substructures that are significant in a GRN under gene duplication.\end{abstract}

\pagenumbering{arabic} 
\label{chapter:intro}

\label{chapter:motifs}
\section{Introduction}

Traditionally the study of gene duplication has a primary focus on sequence duplication, which involves discovering similar regions of DNA that contain a homologous gene \citep{zhang1914}. Of equal importance is how the regulation evolves during the duplication process. This evolution of function has yet to be studied in the same depth as duplication of sequence. In this paper we will use combinatorial probability to investigate the duplication of gene regulation inside of a genetic regulatory network (GRN).
 
GRNs are collections of genes and their products that interact with one another to control a specific cell function and they play a vital role in different cellular processes. These genes and their regulatory elements create complex networks with unexplained design properties \citep{milo2}. An approach to discovering some of the structural design properties is to look for network motifs, which are defined as patterns of interconnections that recur more often in the complex network than in a special randomized network \citep{milo2}.  These motifs can be found in complex networks, but in this paper we will consider motifs within eukaryotic transcriptional networks \citep{lee2002} \citep{Harbison2004}.

Network motifs in transcriptional networks constitute the building blocks of these networks \citep{milo2} \citep{milo3}.  They can be constructed by identifying most or all of the transcription factors in a genome and identifying their binding sites to other genes in the genome \citep{Ren2000}.  The result is that the transcription factors with the binding sites link genes into a transcriptional network for an entire eukaryotic genome \citep{lee2002}  \citep{Harbison2004}.  A key challenge is to identify these network motifs within a eukaryotic transcriptional network.  The usual tool for doing so is simulation and envisioning that instances of a particular motif as the product of randomized networks of similar structure\citep{milo2}. However in this paper we  examine the evolution of gene-family-specific network motifs, which will be referred to as subnetwork motifs, under gene duplication. 

\begin{figure}[h]
\centering
	\includegraphics[width=50mm]{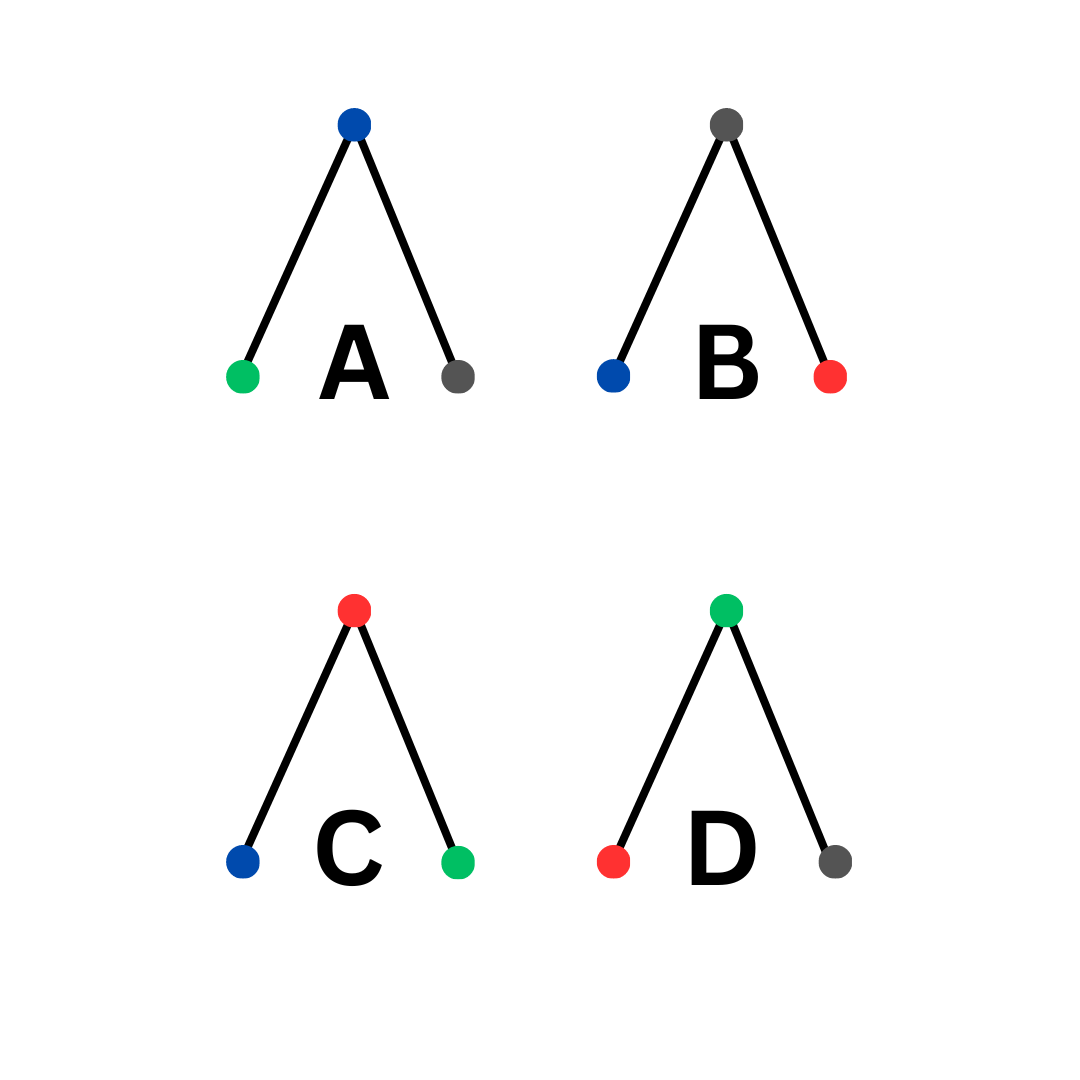}
	\caption[Subnetwork Motifs vs. Network Motfs]{Each color (blue, green, red, grey) represents a different gene family. Then under the definition of network motifs found in \citep{milo2} A, B, C, D would be grouped under 3-node network motifs, since they are isomorphic subgraphs. However, under the subnetwork motifs definition presented in this thesis A, B, C, D would be considered 4 different subnetwork motifs, since the specified gene families are different in each subgraph.}
	\label{fig:milo_vs_sub}
\end{figure}

Subnetwork motifs are distinguished from network motifs by being specific substructures associated with particular gene families, whereas network motifs represent substructures that are isomorphic across different gene families. The difference can also been seen in Figure \ref{fig:milo_vs_sub}. While network motifs are applicable to complex networks in general, this paper aims to further explore the concept of network motifs by focusing on the occurrences of subnetwork motifs in a GRN. By focusing on subnetwork motifs found in GRNs, this research intends to help researchers determine which gene families and regulation relationship are significant enough to explore. 

The framework of subnetwork motifs for identifying building blocks assumes that there are functional labels on the families in the GRN.  This is not the case for network motifs \citep{milo2}. The discovery of these important subnetwork motifs will depend on the how the regulation matrix was inferred and what data was used to infer it.  For example, in the case of the \textit{Arabidopsis} clock, a feedforward network motif was fruitfully identified \citep{miller}.  On the other hand a transcriptional network for yeast was not fruitful in identifying some of the regulatory links (particularly post-transcriptional links) in several well studied GRNs \citep{mazurie}. What is also interesting about subnetwork motifs is that they incorporate evolutionary information, which network motifs do not do \citep{wagner}. This can be used to further validate subnetwork motifs as over or under-represented. Also, some healthy skepticism should be maintained when applying subnetwork motifs to identifying important building blocks that are based on both the data specifying the regulatory links and the evolutionary information available on the gene families in the motif.

Identifying subnetwork motifs in GRNs gives clues to the function of genetic networks. Network motifs in GRNs can help us to understand the functions of networks and allow us to compare how the regulation of networks has evolved or can be evolved by engineering \citep{fran}. 
At the heart of synthetic biology is altering regulation rather than sequence, and it is now possible to create libraries of regulators with different functions and carry out directed evolution on the wiring of genetic networks for improved enzyme activity \citep{fran}.  In order for this program of directed evolution through the regulation to work, it is necessary to identify “the selected motifs” driving the evolution of new functions. Being able to identify “significant” network motifs is central to identifying the function of regulatory networks from their components, how they evolve, and how synthetic biology can be used in protein engineering through selection on regulation rather than sequence. The results of this paper will provide a way to identify the “significant” network motifs.  

The gene duplication and inheritance model that is presented in this paper is adapted from a gene duplication model for biological networks presented in \citep{fanchung}. In the model presented in \citep{fanchung}, each inheritance mode, Full Duplication and Partial Duplication, are controlled by a single probability respectively. However, in our model each gene family has its own probability of inheriting regulatory links through duplication. That is to say that the gene duplication and inheritance model uses a vector of probabilities to determine the inheritance of regulatory relationships and the vector is determined by the mode of inheritance. In the end, we find that our model is a generalization of the model presented in \citep{fanchung}. 

In this paper we will use combinatorial probability to study the occurrences of subnetwork motifs in our gene duplication and inheritance model. We begin by defining a stochastic gene duplication process that governs the gene duplication and inheritance model under all variations of the inheritance vector. Then we define in detail a subnetwork motif and observe their occurrences in both Full and Partial Duplication. 

To create the framework for carrying out significance tests for subnetwork motifs we will calculate two moments: the mean and the variance. We will begin with defining the gene duplication process that applies to both Full Duplication and Partial Duplication. Then we will differentiate the models of inheritance before calculating the moments for each model. In the end we will present exact results and some asymptotic results for the moments.

\section{Constructing the Gene Duplication and Inheritance Model}

\subsection{The Gene Duplication Process}

The $\textit{gene}$ $\textit{duplication}$ $\textit{process}$ is a stochastic process that begins with $m$ individual genes, where $n \geq m \geq 1.$ Initially, each of these genes is the sole member of its gene family, called the $i^{th}$ family for $i=1,2,3, ... , m$ in some arbitrary but fixed order. At each step, a random gene is selected to be duplicated. If a gene in the $i^{th}$ family is duplicated then the new duplicated gene belongs to the $i^{th}$ family. After $d$ duplications the total number of genes will be $n = m+d.$ Often $n$ will be referred to as the $\it{stage}$ of the gene duplication process. This gene duplication model 

\begin{proposition}\label{prop_2.1}
Suppose there are $c_{i}$ genes in the $i^{th}$ family, where $c_{i} \in \mathbb{Z}^+$ for $i = 1, ...,m$. Then $\vec{c} = (c_{1},c_{2}, ... , c_{m})$ is a \textit{composition} of $n$ into $m$ parts where $\sum {c}_{i} = n$. From now on we will discuss the results of a series of  duplications in terms of compositions.  It turns out that for given $m$ and $n$ all such compositions are equally likely. 
\end{proposition}
This result characterizes the uniformity of the duplication process that governs both Full and Partial Duplication inheritance modes.
\begin{proof}
We induct on $n \geq m$. For the base case $n=m$ the only possible composition is $(1, ... ,1)$. Since $ \binom{m-1}{m-1} = 1$ the base case is verified.  

For the induction step, assume that that $n > m$, and let $p$ be the probability that after $n-m$ duplications the composition is $(c_{1}, ..., c_{m}).$ Here $c_{i} \geq 1$ for $i = 1, ..., m$ and $c_{1} + \cdots + c_{m} = n$. Now $p=p_{1} + \cdots + p_{m}$, where $p_{i}$ is the probability that the given composition is the result of duplicating a member of the $i^{th}$ family at the $(n-m)^{th}$ duplication step. In order for the $i^{th}$ family to be duplicated at the $(n-m)^{th}$ step the composition after the previous step must have been $\vec{\gamma} = (c_{1}, ..., c_{i-1}, c_{i} - 1, c_{i+1}, ..., c_{m})$. We claim that $p_{i} = (c_{i} - 1)/\binom{n-2}{m-1}$. If $c_{i} = 1$ this yields $p_{i} = 0$, which is correct since $\vec{\gamma}$ is not a possible composition in this case. If $c_{i} \geq 2$ then the probability of $\vec{\gamma}$ is $1/\binom{n-2}{m-1}$ by the induction hypothesis, and the probability given $\vec{\gamma}$ that the next duplication is in the $i^{th}$ family is $(c_{i}-1)/(n-1)$. Taking the product we find 
\begin{center} $p_{i} = (c_{i} - 1)/ \left( (n-1)\binom{n-2}{m-1} \right)$. \end{center} Since \begin {center} $(n-1) \binom{n-2}{m-1} = (n-m) \binom{n-1}{m-1}$ \end{center}  and \begin{center} $\sum_{i-1}^{m} (c_{i} -1) = n - m $ \end{center} we have that \begin{center} $ p_ = \sum_{i=1}^{m} p_{i} = (n-m) / \left((n-m) \binom{n-1}{m-1} \right) = 1 / \binom{n-1}{m-1}$. \end{center}

Finally, if we sum over all the possible compositions of $n$ into $m$ parts, then the total probability must be 1, so there are $\binom{n-1}{m-1}$ equally likely compositions.
\end{proof}

It is important to note that there are multiple ways to arrive at $\binom{n-1}{m-1}$ as the total number of equally likely compositions of $n$ into $m$ parts. From a combinatorial point of view let a family be a bin for gene markers and each family is understood to initially contain $1$ gene marker which will not be shown explicitly, leaving $n-m$ gene markers represented by $0$, and $m-1$ dividers represented by $|$, to be arranged in linear order. Note that the combinatorial symbols for gene markers are identical and the dividers are identical. Then the linear arrangement has $n-1$ locations that have $m-1$ dividers.

Consider the case where $m = 3$ and $n= 6$ such that $\vec{c} = (2,1,3).$ Then the combinatorial representation is $$0 | | 0 0$$ \noindent where the two dividers creates three blocks of zeros with lengths $1, 0,$ and $2$ representing family sizes $2, 1,$ and $3$ respectively.

The number of gene markers and dividers that are going to be arranged in linear order is the number of dividers plus the additional gene markers that need to be place and $m- 1 + n-m = n-1$. Thus there are $\binom{n-1}{m-1}$ possible linear arrangements for $n$ gene markers to be placed in $m$ families.

An ordinary generating function is an alternative way of counting the compositions of $n$ into $m$ parts. Using a generating function an infinite sequence of numbers can be expressed by allowing those numbers to be the coefficients of a formal power series \citep{concretemath}. We find that the use of generating functions will be a convenient way to calculate some of the numbers we are going to need when analyzing subnetwork motifs since generating functions can be represented as Taylor Series of explicit rational functions. 

If you let an infinite series of numbers be denoted by $g_{0}, g_{1}, ..., g_{j}$ where $g_{i} \geq 0$ then its \emph{generating function} is defined as the infinite series $$g(x) = (g_{0}x^{0} +g_{1} x^{1} + ... + g_{j}x^{j} + ...) = (g_{0} +g_{1} x + ... + g_{j}x^{j} + ...).$$

\noindent Let $\alpha$ be a real number and $j \geq 0$ be an integer then 
$$\binom{\alpha}{j} \equiv \frac{(\alpha)_{j}}{j!}$$ 
where $(\alpha)_{j}$ is a falling factorial.
Note that this is an extension of the standard binomial definition since taking $\alpha$ to be a non-negative integer would yield the standard binomial coefficient. 

\begin{lemma}\label{genbinom}
Let $\alpha \geq 0$ \textrm{ be a real number and } $j \geq 0$ be an integer. Then,

$$[x^{j}](1-x)^{-\alpha} = \binom{\alpha + j -1}{j}.$$
\end{lemma}

\begin{proof}
This follows from the Taylor Series at $x=0$. Note that for any integer $j \geq 0$, the operator $[x^{j}]$ returns the coefficient of $x^{j}$ in a power series. Thus $[x^{j}] g(x) = g_{j}$ in the infinite power series. Notice that $$[x^{j}] xg(x) = 0$$ \noindent if $j=0$ \noindent and $$[x^{j}] xg(x) = g_{j-1}$$ \noindent if $$j \geq 1.$$ \noindent In general, $$[x^{j}] x^{\ell}g(x) = 0$$ \noindent if $j < \ell$ and $$[x^{j}] x^{\ell}g(x) = g_{j-\ell}$$ \noindent if $j \geq \ell$.
\end{proof}

\begin{lemma}\label{OGF}
Let $\ell \geq 0$ be a non-negative integer. Then the ordinary generating function for the number of compositions into $\ell$ parts is $$x^{\ell} (1-x)^{-\ell}.$$
\end{lemma}

\begin{proof}
When $\ell \geq 1$ the Lemma follows directly Lemma \ref{genbinom} and Proposition \ref{prop_2.1}. However, when $\ell = 0$ the Lemma still stands since there is only 1 composition of $0 \textrm{ into } 0$ parts and there are no compositions of $n$ into 0 parts when $n \geq 0$.
\end{proof}

There is also an alternative derivation of the generating function that we will find useful. For compositions of $n$ into 1 part we know the generating function to be \\$g(x) = (x + x^{2} + x^{3} + ...) = \frac{x}{1-x}.$ In order to obtain the generating function for the number of compositions of $n$ into $\ell$ parts we can allow each of the $\ell$ parts be denoted by $g(x)$. Therefore the generating function for the compositions of $n$ into $\ell$ parts is 

$$g(x)^{\ell} = (x + x^{2} + x^{3} + ...)^{\ell} = (\frac{x}{1-x})^{\ell}.$$

Let $m \geq 1$ be fixed with $n \geq m$, and consider the process of starting from the composition $(1, ..., 1)$ (of dimension $m$) and performing $n-m$ duplications. The resulting vector $(X_{m,n}^{(1)}, ..., X_{m,n}^{(m)})$ is a composition of $n$ into $m$ parts. From Proposition \ref{prop_2.1} we know that there are $\binom{n-1}{m-1}$ such vectors, all with the same probability of occurring. 

We will find the following special notation for vectors helpful in the remainder of the thesis. For vectors $\vec{x} = (x_{1}, ... , x_{h})$ and $\vec{y} = (y_{1}, ... , y_{h})$ of the same dimension $h$ we let $\vec{y} \leq \vec{x}$ denote the conjunction of the $h$ inequalities ${y}_{1} \leq {x}_{1}, ... , {y}_{h} \leq {x}_{h}.$ We define $<, \geq,$ and $>$ for vectors similarly. We also define a special operator $ ||\textrm{ }\vec{x}\textrm{ }|| $ such that $||\textrm{ }\vec{x}\textrm{ }||= \sum x_{i}$, and special constant vectors $\vec{0} = (0, ..., 0)$ and $\vec{1} = (1, ... ,1)$. For the latter the dimension is to be made clear by context. 

In general, the gene duplication process should be seen as a random markov process. To make calculating the expectations more convenient we present the following lemma. 

\begin{lemma}\label{exp_lemma} Let $m \text{ and } n$ be integers such that $1 \leq m \leq n$, and for $\nolinebreak{i = 1, ... , m}$ let $\mathbb{U}_{i}$ be a real valued function over the positive integers and $\nolinebreak{\mathbb{U}_{i}(x) = \sum_{j=1}^{\infty} U_{i}(j)x^{j}}.$ Then the sum of $\prod_{i=1}^{m}U_{i}(c_{i})$ over the compositions $\vec{c} = (c_{1}, ..., c_{m})$ of $n$ into $m$ parts is $$\sum_{\vec{c}} \Big(\prod_{i=1}^{m} U_{i}(c_{i})\Big) = [x^{n}]\prod_{i=1}^{m} \mathbb{U}_{i}(x).$$
\end{lemma}

\begin{proof}
Suppose $ (c_{1}, ..., c_{m})$ is a partition of $n$ into $m$ parts. Then $\prod_{i=1}^{m}U_{i}(c_{i})$ arises as $[x^{n}]({U}_{1}(c_{1})x^{c_{1}} \cdot ... \cdot {U}_{m}(c_{m})x^{c_{m}})$ since $c_{1} + ... + c_{m} = n$. That is one monomial that is obtained from expanding the product into a single generating function. Each partition of $n$ into $m$ parts similarly contributes its own share to $[x^{n}]\prod_{i=1}^{m}\mathbb{U}_{i}(x)$. 

To obtain the result of the Lemma we expand the product of these generating functions into a sum of monomials. Then to arrive at a single generating function, we collect the monomials that correspond to each power of $x$. Note that since the exponents sum to $n$,  the only monomials that can correspond to $x^{n}$ must have exponents that sum to $n$ and therefore arise as one of the partitions of $n$ into $m$ parts.
\end{proof}

\section{Introduction to Subnetwork Motifs}\label{intro_to_motifs}

The subnetwork motifs discussed in this thesis are gene-family-specific network motifs. A network motif is said to be a pattern of interconnections that occur more often in a complex network than a special random network \citep{milo2}. From a biological point of view, a subnetwork motif is a gene-family-specific network motif that occurs more in real data than expected from the moments calculated in the coming sections. It is important to note that the gene-family-specific-substructures discussed in this thesis will be called $\it{subnetwork\textrm{ }motifs}$ although for biological applications $\it{subnetwork\textrm{ }motif\textrm{ }candidate}$ would be more apposite. 

Subnetwork motifs are going to be built from the genes that arise from the gene duplication process. For the purposes of this thesis, a subnetwork motif $\mathbb{M}$ is characterized by $k$ gene families and the chances of creating a new subnetwork motif instance in duplication. Given a subnetwork motif $\mathbb{M}$, we will assume the indices of the families that belong to $\mathbb{M}$ are $1$ to $k$ for notational convenience. Then by definition the set of $k$ original genes forms the original instance of $\mathbb{M}$. At any stage $n$ during the duplication process the set of instances of $\mathbb{M}$ will be denoted $\mathbb{M}(n)$. When $n=m$, the only subnetwork motif instance present is the original instance of $\mathbb{M}$; therefore $|\mathbb{M}(m)| = 1$. Since instances of $\mathbb{M}$ will always have dimension $k$ we will denote the vector of family sizes that belong to $\mathbb{M}$ as $\vec{s} = (s_{1}, ..., s_{k})$, where $s_{i} = |X^{(i)}_{m,n}|$. For any $n \geq m$ new instances of the subnetwork motif are possible at stage $n +1.$ Suppose there is a subnetwork motif instance $\mathscr{I} = (a_{1}, ..., a_{i}, ... a_{k}) \in \mathbb{M}(n)$. If a gene $a_{i}^{\prime}$ is duplicated from $a_{i}$ at stage $n+1\textrm{ then }\mathscr{I}^{\prime} =(a_{1}, ..., a_{i}^{\prime}, ... a_{k})$ is a potential new instance of $\mathbb{M}$ and if $\mathscr{I}^{\prime}$ is inherited then $\mathscr{I}^{\prime} \in \mathbb{M}(n+1).$ At any stage $n\geq m,\textrm{ }\mathbb{M}(n)$ consists of the original instance of $\mathbb{M}$ and the additional instances of $\mathbb{M}^{\prime}$ that were inherited through the duplication process.

Each subnetwork motif $\mathbb{M}$ has an associated vector of probabilities, $\vec{\pi}=(\pi_1, ..., \pi_k)$. Here $\pi_i$ is the probability that an instance $\mathscr{I} =(a_{1}, ..., a_{i}, ... a_{k}) \in \mathbb{M}(n)$ gives rise by inheritance to the new instance $\mathscr{I}^{\prime} =(a_{1}, ..., a_{i}^{\prime}, ... a_{k})$ when the gene in the $i^{th}$ family is duplicated. The general concept of subnetwork motif inheritance is central to the thesis. Simpler results are obtained when $\vec{\pi} = \vec{1}$ which is called Full Duplication. A more general case is also analyzed where $\vec{0} \leq \vec{\pi} \leq \vec{1}$ and this is called Partial Duplication. 

To study the expected size of $\mathbb{M}(n),$ the gene duplication process and the subnetwork motif inheritances process are combined into one process. The $\textit{random}\textrm{ }\textit{duplication}\textrm{ }\textit{and}\textrm{ }\textit{inheritance} \textit{ } \textit{process}$ is a stochastic process that begins with $m$ individual genes and a $k$ sized subnetwork motif that has an associated vector of probabilities $\vec{\pi} = (\pi_{1}, ..., \pi_{k})$, where $n \geq m \geq k \geq 1$ and $\vec{0} \leq \vec{\pi} \leq \vec{1}$. Initially, each of these genes are the sole member of its gene family called the $i^{th}$ family and $\mathbb{M}$ is the original subnetwork motif instance in $\mathbb{M}(n)$. At each step, a random gene is selected for duplication. If a gene is duplicated from the $i^{th}$ family, then the new gene belongs to the $i^{th}$ family and there is a $\pi_{i}$ chance that the regulations are inherited at that step if $1 \leq i \leq k$. After $n-m$ duplications we are interested in the size of $\mathbb{M}(n)$. 

\subsection{Full Duplication}

This section will cover the $\it{Full\textrm{ }Duplication}$ model, which is the inheritance model that is controlled by the probability vector $\vec{\pi} = \vec{1}$. Suppose $k \geq1$ and $\mathbb{M}$ is a subnetwork motif of size $k$. Every possible instance is duplicated since $\vec{\pi} = \vec{1}$.  Therefore $\mathbb{M}(n) = X_{m,n}^{(1)} \times ... \times X_{m,n}^{(k)}$, so that $|\mathbb{M}(n)| = |X^{(1)}_{m,n}|  \cdot ... \cdot  |X^{(k)}_{m,n}|$. We will start by analyzing the expected value of $|\mathbb{M}(n)|$ given $m,n,$ and $k$.


\begin{theorem}\label{full_dup_fixed_n_1st_mom}
Assume $\mathbb{M}$ is a subnetwork motif of size $k$ and $1 \leq k \leq m \leq n.$ Then the expected number of instances of $\mathbb{M}$ given the random duplication process is
$$\mathbb{E}\Big(|\mathbb{M}(n)| \textrm{ } ; \textrm{ } k,m,n\Big) = \frac{\Gamma(n+k)\Gamma(m)}{\Gamma(n)\Gamma(m+k)}.$$
\end{theorem}
This result allows us to construct a significance test for subnetwork motifs using the mean of the number of instances of $\mathbb{M}$ under Full Duplication. 
\begin{proof}
Let $\vec{c} = (c_{1}, ..., c_{m})$ where $c_{i}$ is the size of the $i^{th}$ family, that is $c_{i} = |X^{(i)}_{m,n}|$. Then $|\mathbb{M}(n)| = c_{1} \cdot ... \cdot c_{k} = c_{1} \cdot ... \cdot c_{k} \cdot 1^{m-k}$. To calculate the expectation we use Lemma \ref{exp_lemma} to evaluate  $\sum |\mathbb{M}(n)|$ over all the compositions of $n$ into $m$ parts then divide the result by $\binom{n-1}{m-1}. \textrm{ }$ In order apply Lemma \ref{exp_lemma} we let $\mathbb{U}_{i}(x) = (x + 2x^{2} + ... )$ when $1 \leq i \leq k$ and $\mathbb{U}_{i}(x) = (x + x^{2} + ... )$ when $k+1 \leq i \leq m$. We will call the series $a(x)$ and $b(x)$ for convenience. When $1 \leq i \leq k$ the $i^{th}$ family contributes to $|\mathbb{M}(n)|$ and $a(x) =  (x + 2x^{2} + ... ) = \frac{x}{(1-x)^{2}}$. Furthermore, when $k+1 \leq i \leq m$ the $i^{th}$ family does not contribute to $|\mathbb{M}(n)|$ and $b(x) = (x + x^{2} + ... ) = \frac{x}{(1-x)}$.

It is easy to see the direct correlation of the above generating functions as rational functions of $x$. However, it can also be seen by applying Lemma \ref{genbinom} and its proof when $\alpha = 2$ for $a(x)$ and $\alpha = 1$ for $b(x)$. Thus, applying Lemma \ref{exp_lemma} with $\mathbb{U}_{i}(x) = a(x)$ if $k+1 \leq i \leq m$ and $\mathbb{U}_{i}(x) = b(x)$ if $1 \leq i \leq k$ we obtain 

\begin{align*}
\sum_{\vec{c}} |\mathbb{M}(n)| &= x^{[n]}\big(\frac{x}{1-x}\big)^{m-k}\big(\frac{x}{(1-x)^{2}}\big)^{k} \\
&= x^{[n]} \big(\frac{x^{m}}{(1-x)^{m+k}}\big) \\
&= \binom{n+k-1}{m+k-1},
\end{align*}
\noindent where the third equality follows from Lemma \ref{genbinom} when $\alpha = m+k$. 

Note that the result of the Theorem is expressed in terms of the gamma function as it will be convenient in the Partial Duplication section. The gamma function $\Gamma(z) = (z-1)!$ is defined on all of the complex plane except for $z \leq 0$ an integer \citep{mathfun}. Therefore, $\Gamma(n) = (n-1)!$ as long as $n \geq 1$ is an integer.

Recall from Proposition \ref{prop_2.1} that there are exactly $\binom{n-1}{m-1}$ compositions of $n$ into $m$ parts  and they are equally likely. Thus

\begin{align*}
\mathbb{E}\Big(|\mathbb{M}(n)| \textrm{ } ; \textrm{ } k,m,n\Big) &= \frac{\binom{n+k-1}{m+k-1}}{\binom{n-1}{m-1}} \\
&= \frac{\Gamma(n+k)\Gamma(m)}{\Gamma(n)\Gamma(m+k)}.
\end{align*}
\end{proof}

Similar to the combinatorial explanation after Proposition \ref{prop_2.1}, $\binom{n+k-1}{m+k-1}$ can also be derived combinatorially. We are now interested in the linear arrangements considering $k$ sized subnetwork motifs. Given a particular $\vec{c} = (c_{1}, ..., c_{k}, c_{k+1}, ..., c_{m})$, we add a selector to the first $k$ families such that there are $k$ selectors represented by $|$. The selectors indicate which gene in that family has been selected for the subnetwork motif so that the $k$ selected genes form a subnetwork motif. This leaves $n-m$ gene markers represented by $0$ and $m+k-1$ selectors and dividers represented by $|$ to be arranged in linear order. Note that the first $k$ odd placements for pipes are selectors and the first $k$ even placements for pipes are dividers. 

Consider the case where $k=2, m = 3$ and $n= 6$ such that $\vec{c} = (3,1,2)$. Let $g_{i}$ be the original gene for the $i^{th}$ family for $1 \leq i \leq m$ and $g_{i}^{\prime}$ be the first duplicated gene in the $i^{th}$ family where the number of primes represents the order in which the gene was duplicated. Then the combinatorial representation for $(g_{1}^{\prime \prime}, g_{2}, g_{3}^{\prime})$ is $$0 0 | | | | 0$$ \noindent where the first pipe is selecting the second duplicated gene marker for the first family, the second pipe is the first divider, the third pipe is selecting the original gene marker from the second family, and the fourth pipe is the second divider. 

The number of genes, dividers, and selectors that are going to be arranged in linear order is the number of dividers and selectors plus the number of additional genes that need to be placed and $m + k - 1 + n - m = n+k-1$. Thus the total number of linear arrangements considering a $k$ sized subnetwork motifs is  $\binom{n+k-1}{m+k-1}$. 

\begin{corollary}\label{bound_exp_fixedm}
Suppose $1 \leq k \leq m \leq n$ and $n \geq k^{2}.$ Then the expected number of instances of subnetwork motifs of size $k$ in Full Duplication satisfies 

$$\frac{n^{k}}{(m+k-1)_{k}} \leq \mathbb{E}\Big( |\mathbb{M}(n)| \textrm{ } ; \textrm{ } k,m,n\Big) \leq \frac{n^{k}}{(m+k-1)_{k}} \Big(1+\frac{3k^{2}}{4n}\Big).$$
\end{corollary}

\begin{proof}
From Theorem \ref{full_dup_fixed_n_1st_mom} we know that $\mathbb{E}\Big(|\mathbb{M}(n)| \textrm{ } ; \textrm{ } k,m,n\Big)$ is

$$\frac{\Gamma(n+k)\Gamma(m)}{\Gamma(n)\Gamma(m+k)} = \frac{(n+k-1)_{k}}{(m+k-1)_{k}}.$$

The bounds for the expectation fall directly from Lemma \ref{lemma3} (occurs in Section \ref{chapter:formulae}).
\end{proof}

We are going to use the same general approach to evaluating the second moment of the number of instances of $\mathbb{M}$ in Full Duplication which we will denote $\mathbb{E}\Big (|\mathbb{M}(n)|^{2} \textrm{ };\textrm{ } k,m,n \Big).$

\begin{theorem}\label{full_dup_fixed_n_2nd_mom}
Assume $\mathbb{M}$ is a subnetwork motif of size $k$ and $1 \leq k \leq m \leq n$. Then the second moment of the number of instances of $\mathbb{M}$ in Full Duplication is evaluates to $$\frac{\Gamma(n+k)\Gamma(m)\Gamma(n-m+1)}{\Gamma(n)} \mathlarger{\mathlarger{\sum}}_{i=0}^{k} \binom{k}{i} 2^{i} \big(\Gamma(m+k+i)\Gamma(n-m-i+1)\big)^{-1}.$$ 
\end{theorem}
This result allows us to calculate the variance of the number of instances of $\mathbb{M}$ under Full Duplication, which would be required for a significance test. 
\begin{proof}
Let $\vec{c} = (c_{1}, ..., c_{m})$ where $c_{i}$ is the size of the $i^{th}$ family, that is $c_{i} = |X^{(i)}_{m,n}|$. Thus $|\mathbb{M}(n)|^{2} =( c_{1}\cdot ... \cdot c_{k})^{2} = (c_{1}^{2} \cdot ... \cdot c_{k}^{2}) \cdot 1^{m-k}$. In order to calculate the second moment we can proceed as in the proof of Theorem $\ref{full_dup_fixed_n_1st_mom}$ and use Lemma \ref{exp_lemma} to evaluate $\sum |\mathbb{M}(n)|^{2}$ over all the compositions of $n$ into $m$ parts then divide the result by $\binom{n-1}{m-1}$. Recall from the proof of Theorem $\ref{full_dup_fixed_n_1st_mom}$ the ordinary generating functions used to evaluate $\sum \mathbb{M}(n)$ over the compositions of $n$ into $m$ parts are:
$$a(x) = (x  + 2x^{2} + ... )= \Big(\frac{x}{(1-x)^{2}}\Big) \textrm{ and } b(x) = (x + x^{2} + x^{3} + ...)= \Big(\frac{x}{1-x}\Big).$$
In order to apply Lemma \ref{exp_lemma} we must modify $a(x)$ since the product first $k$ family sizes has been squared. To evaluate the second moment we will replace $a(x)$ with another series we denote $y(x)$ where 
$$y(x) = x \cdot \frac{d}{dx}\big(a(x)\big) = (1 + 4x + 9x^{2}  + ...).$$
When $1 \leq i \leq k$ the $i^{th}$ family contributes to $|\mathbb{M}(n)|$, $y(x) =  \frac{2x^{2}}{(1-x)^{3}} +  \frac{x}{(1-x)^{2}}$ and when $k+1 \leq i \leq m$ the family does not contribute to $|\mathbb{M}(n)|$, $b(x) = \frac{x}{1-x}$. Therefore, we apply Lemma \ref{exp_lemma} with $\mathbb{U}_{i}(x) = y(x)$ when $1 \leq i \leq k$ and $\mathbb{U}_{i}(x) = b(x)\textrm{ when }k+1 \leq i$. By the binomial theorem $y(x)^{k}$ can be expressed as
$$y(x)^{k} = \sum_{i=0}^{k} \binom{k}{i}  \Big(\frac{2x^{2}}{(1-x)^{3}}\Big)^{i}\Big(\frac{x}{(1-x)^{2}}\Big)^{k-i},$$ 
\noindent which gives the following
\begin{align*}
\sum_{\vec{c}} |\mathbb{M}(n)|^{2} &= x^{[n]}(\frac{x}{1-x})^{m-k}\sum_{i=0}^{k} \binom{k}{i} (\frac{2x^{2}}{(1-x)^{3}})^{i}(\frac{x}{(1-x)^{2}})^{k-i} \\
&= [x^{n}]\sum_{i=0}^{k} \binom{k}{i} \frac{2^{i}x^{i}x^{m}}{(1-x)^{k+m+i}} \\
&= \mathlarger{\mathlarger{\sum}}_{i=0}^{k} 2^{i} \binom{k}{i} \binom{n+k-1}{n-m-i} \\
&= \mathlarger{\mathlarger{\sum}}_{i=0}^{k} 2^{i}\binom{k}{i} \binom {n+k-1}{m-1+k+i}.
\end{align*}
Since the $\binom{n-1}{m-1}$ compositions of $n$ into $m$ parts are equally likely 
$$\mathbb{E}\Big (|\mathbb{M}(n)|^{2} \textrm{ };\textrm{ } k,m,n \Big) = \binom{n-1}{m-1}^{-1} \cdot \mathlarger{\mathlarger{\sum}}_{i=0}^{k} 2^{i}\binom{k}{i} \binom {n+k-1}{m-1+k+i},$$
which evaluates to 
$$\frac{\Gamma(n+k)\Gamma(m)\Gamma(n-m+1)}{\Gamma(n)} \mathlarger{\mathlarger{\sum}}_{i=0}^{k} 2^{i}\binom{k}{i} \big(\Gamma(m+k+i)\Gamma(n-m-i+1)\big)^{-1},$$
\noindent where $\Gamma(z)$ is the special function that is noted in the proof of Theorem \ref{full_dup_fixed_n_1st_mom}.
\end{proof}

Similar to Theorem \ref{full_dup_fixed_n_1st_mom}, we will find a combinatorial explanation for the binomial coefficient portion of the second moment calculation. Consider an ordered pair of subnetwork motifs $\nolinebreak{(\mathscr{I}_{1}, \mathscr{I}_{2}) \in \mathbb{M}(n)^{2}}$ where $\mathscr{I}_{1}$ is inherited first and $\mathscr{I}_{2}$ is inherited second. Since we are observing ordered pairs of subnetwork motifs there will be $0 \leq i \leq k$ families that have two gene indicators selecting for two different gene markers and $k-i$ families that will have one gene indicator selecting for the same gene marker. Consider the case where $k=3, m=4, n=6,$ and $i=2$ where $\vec{c} = (2,2,1,1).$ Suppose the first two families have two gene indicators, $\mathscr{I}_{1} = (g_{1}, g_{2}, g_{3})$ comes first, $\mathscr{I}_{2} = (g_{1}^{\prime}, g_{2}^{\prime}, g_{3})$ come second, $g_{1}$ is duplicated earlier, and $g_{2}$ is duplicated later. Then the combinatorial representation of the ordered pair of subnetwork motifs is $$| 0 | | | 0 | | | |.$$ \noindent To visually show the distinction between the two types of selectors and dividers we show an annotated version of the combinatorial representation $$\underline{|} 0 \underline{|} | \underline{|} 0 \underline{|} | \underline{|} |$$\noindent where $\underline{|}$ represents selectors that select for different genes and $\hat{|}$ represents selectors that select for the same gene.

Given an ordered pair of subnetwork motifs, the process begins with $m$ original genes that are not shown explicitly, there are $n-m$ additional gene markers that need to be placed. Note that for each of the $i$ families with double gene selectors there is a gene indicator that is not shown explicitly to ensure separation for each pair of selectors. Therefore, there are  $m-1+k+i$ dividers and selectors and the number of gene markers, dividers, and selectors that are going to be arranged in linear order is the number of dividers and selectors plus the number of additional gene markers and $m-1 + k+i + n-m+i = n+k-1.$ Thus the linear arrangement has $n+k-1$ locations that have $m-1+k+i$ selectors and dividers. 

Note that there are $\binom{k}{i}$ ways the families with 2 selectors could have been chosen and $2^{i}$ ways the selectors could be arranged, since the selector that is selecting for the first subnetwork motif can come before or after the selector that is selecting for the second subnetwork motif. 

\begin{corollary}\label{fd_variance}
Suppose $1 \leq k \leq m \leq n.$ Then the variance of the number of instances of subnetwork motifs of size $k$ in Full Duplication is 

$$ \binom{n-1}{m-1}^{-1} \cdot \Big( \mathlarger{\mathlarger{\sum}}_{i=0}^{k} \binom{k}{i} 2^{i} \binom {n+k-1}{m+k-1+i}\Big) - \Big(\frac{(n+k-1)_{k}}{(m+k-1)_{k}}\Big)^{2}.$$
\end{corollary}

This result allows us to construct a significance test for subnetwork motifs under Full Duplication. 

\begin{proof}
The corollary follows immediately from Theorems $\ref{full_dup_fixed_n_1st_mom}$ and $\ref{full_dup_fixed_n_2nd_mom}$. Furthermore, when $m=1$ or when $m=n$ the variance is $0$ and non-zero in all other cases. 
\end{proof}

For the rest of the section we will present the inequalities in the following form $$\ineql{n}{rk}{\mathbb{C}}{\mathbb{L}}{i} \leq \mathbb{Y} \leq \inequ{n}{rk}{\mathbb{C}}{\mathbb{U}}{i}$$ \noindent where the power of $n$ depends on the $r^{th}$ moment of $\mathbb{Y}, \mathbb{C}$ is the constant for the main term and $\mathbb{L}$ and $\mathbb{U}$ are the error terms for the upper and lower bounds respectively. Using this notation to present Corollary \ref{bound_exp_fixedm} the inequalities will take the following form $$\ineql{n}{k}{\mathbb{C}}{\mathbb{L}}{1} \leq \mathbb{E}\Big(|\mathbb{M}(n)| \textrm{ } ; \textrm{ } k,m,n\Big) \leq \inequ{n}{k}{\mathbb{C}}{\mathbb{U}}{1}$$ \noindent where $\mathbb{C}_{1} = \Gamma{(m)}/\Gamma{(m+k)}, \mathbb{L}_{1} = 0, \textrm{ and }\mathbb{U}_{1} = 3k^{2}$. 

\begin{corollary}\label{fd_sec_mom_bounds}
Suppose $1 \leq k \leq m \leq n$ and $n \geq 3m^{2}$. Then the second moment of the number of instances of subnetwork motifs of size $k$ in Full Duplication is 
$$\ineql{n}{2k}{\mathbb{C}}{\mathbb{L}}{2} \leq \mathbb{E}\Big((|\mathbb{M}(n)^{2}| \textrm{ } ; \textrm{ } k,m,n\Big) \leq \inequ{n}{2k}{\mathbb{C}}{\mathbb{U}}{2}$$ 
where $\mathbb{C}_{2} = 2^{k}/(m+2k-1)_{2k}, \mathbb{L}_{2} = 4m^{2}, \textrm{ and }\mathbb{U}_{2} = 9m^{2}$. 
\end{corollary}

\begin{proof}
From the proof of Theorem \ref{full_dup_fixed_n_2nd_mom} we know that
\begin{align}
\mathbb{E}\Big(|\mathbb{M}(n)^{2}| \textrm{ } ; \textrm{ } k,m,n\Big) &=  \mathlarger{\mathlarger{\sum}}_{i=0}^{k} \binom{k}{i} 2^{i} \binom{n+k-1}{m-1+k+i} \binom{n-1}{m-1}^{-1} \nonumber \\ \nonumber \\
	&=  \mathlarger{\mathlarger{\sum}}_{i=0}^{k} \binom{k}{i} 2^{i} \frac{(n-1+k)!(m-1)!(n-m)!}{(n-1)!(m-1+k+i)!(n-m - i)!} \nonumber \\ \nonumber \\ \label{eqn:label1}
	&=   \mathlarger{\mathlarger{\sum}}_{i=0}^{k} \binom{k}{i} 2^{i} \frac{(n-1+k)_{k}(n-m)_{i}}{(m-1+k+i)_{k+i}}.
\end{align}

In order to obtain bounds for the second moment in Full Duplication we will be bounding Equation (\ref{eqn:label1}) above and below.

Since all the terms in the summation are positive we can take the $i=k$ term as the lower bound since any one of the $k+1$ is a lower bound for the second moment.

Doing this we obtain $$2^{k} \cdot \frac{(n+k-1)_{k}(n-m)_{k}}{(m+2k-1)_{2k}} \leq \mathbb{E}\Big(|\mathbb{M}(n)^2| \textrm{ } ; \textrm{ } k,m,n \Big).$$ In order to derive a simpler lower bound we will bound the term below. Since $m$ and $k$ are both fixed the denominator remains the same and $n^{k}$ is a trivial lower bound for $(n+k-1)_{k}$. Therefore, the only non-trivial bound comes from the term $(n-m)_{k}$. Note that $$(n-m)_{k} = \frac{(n)_{m+k}}{(n)_{m}}$$ which we can bound using Lemma \ref{lemma5} and Lemma \ref{lemma6} (both occur in Section \ref{chapter:formulae}), $$n^{k}(1-\frac{(m+k)^{2}}{n}) \leq \frac{(n)_{m+k}}{(n)_{(m)}}.$$ Therefore the lower bound for the second moment is $$\frac{2^{k}n^{2k}}{(m+2k-1)_{2k}}(1-\frac{4m^{2}}{n}) \leq \mathbb{E}\Big(|\mathbb{M}(n)^{2}| \textrm{ } ; \textrm{ } k,m,n \Big).$$

We will now find the upper bound of the second moment. The upper bound is a summation of the upper bounds of the individual terms. The most important term when bounding the second moment above is the $i=k$ term. We know from the calculation of the lower bound that the $i=k$ term is as follows

$$2^{k} \cdot \frac{(n+k-1)_{k}(n-m)_{k}}{(m+2k-1)_{2k}}.$$

Now we will bound the $i=k$ term above. Since $n^{k}$ is the trivial upper bound for $(n-m)_{k}$ we will only use Lemma \ref{lemma3} (occurs in Section \ref{chapter:formulae}) to bound the term $(n+k-1)_{k}$ above,

$$2^{k} \cdot \frac{(n+k-1)_{k}(n-m)_{k}}{(m+2k-1)_{2k}} \leq \frac{2^{k} \cdot n^{2k}}{(m+2k-1)_{2k}}(1+\frac{3k^{2}}{4n}).$$

It is important to note that each of the $k+1$ terms has the same error bound that can be bounded above
 $$(1 + \frac{3k^{2}}{4n}) \leq \frac{5}{4}$$
since $3k^{2}/4n \leq 1/4.$

When $i \leq k$ let $j = k - 1$ so that $j = 0, 1, ... k$. Then we can bound each of the $k+1$ terms above with the following expression 

\begin{align*}
\mathbb{E}\Big(|\mathbb{M}(n)^{2}| \textrm{ } ; \textrm{ } k,m,n\Big) &\leq \frac{2^{k}n^{2k}}{(m+2k-1)_{2k}} (1 + \frac{3k^{2}}{4n} + \frac{5}{4}(\frac{k(m+2k-1)}{2n} + ... + \frac{(m+2k-2)_{2k}}{2^{k}n^{k}(m+k-1)_{k}})) \\ \\
	&\leq \frac{2^{k}n^{2k}}{(m+2k-1)_{2k}} (1 + \frac{3m^{2}}{4n} + \frac{3m^{2}}{2n} + \frac{5}{4}(\frac{9m^{4}}{8n^{2}} + ... \frac{(3m)^{2m}}{(2n)^{m}(2m)^{m}}))\\ \\
	&\leq \frac{2^{k}n^{2k}}{(m+2k-1)_{2k}} (1 + \frac{3m^{2}}{4n} + \frac{5}{4}\sum_{j=1}^{k} \frac{m^{j}(3m)^{j}}{2^{j}j! n^{j}} ) \\ \\
	&\leq \frac{2^{k}n^{2k}}{(m+2k-1)_{2k}}(1+ \frac{9m^{2}}{2n}).
\end{align*}
\end{proof}

We will now begin to discuss the variance of subnetwork motif occurrences in the Full Duplication mode. Recall that $k \geq 1$ and $\mathbb{M}$ represents a subnetwork motif of size $k$ after $n-m$ duplications. In order to calculate the variance we will use the expressions of the first and second moments. Throughout the rest of the section the variance will be denoted $\mathbb{VAR} \Big( |\mathbb{M}(n)| \textrm{ } ; \textrm{ } k,m,n )\Big)$, where $\vec{s} = (s_{1}, s_{2}, ..., s_{k})$ is the vector of the $k$ family sizes and $\sum \vec{s} = h$.

As usual we will calculate the variance using the second moment and the square of the first moment. In order to simplify the error of the square of the first moment we will find bounds for $e^{2\delta}$ where $\delta$ is as presented in the proof of Lemma \ref{lemma3} (occurs in Section \ref{chapter:formulae}).

It is important to note that the variance is identically 0 when $m=1$ since the number of subnetwork motifs is a point distribution in this case. For $m \geq 2$ we obtain the following inequalities which will be useful for large $n$.

\begin{corollary}\label{fulld_var_bounds}
Let $1 \leq k \leq m \leq n$, $m \geq 2$, and $n \geq 3m^{2}.$ Then the variance of the number of instances of subnetwork motifs of size $k$ in Full Duplication is 
$$\ineql{n}{2k}{\mathbb{C}}{\mathbb{L}}{\nu} \leq \mathbb{VAR} \Big(|\mathbb{M}(n)| \textrm{ } ; \textrm{ } k,m,n )\Big) \leq \inequ{n}{2k}{\mathbb{C}}{\mathbb{U}}{\nu}$$

\noindent where $\mathbb{C}_{\nu} = \mathbb{C}_{2} - \mathbb{C}_{1}^{2}, \mathbb{L}_{\nu} =  (\mathbb{C}_{2}\mathbb{L}_{2}-2\mathbb{C}_{1}^{2}\mathbb{U}_{1})/(\mathbb{C}_{2} -\mathbb{C}_{1}^{2}), \textrm{ and }\\ \mathbb{U}_{\nu} = \mathbb{C}_{2}\mathbb{U}_{2}/(\mathbb{C}_{2} -\mathbb{C}_{1}^{2})$.
\end{corollary}

\begin{proof}
In order to calculate the variance we will take advantage of the inequalities from the first and second moments. Using the standard definition of variance we know 
$$\mathbb{VAR} \Big( |\mathbb{M}(n)| \textrm{ } ; \textrm{ } k,m,n )\Big) = \mathbb{E}\Big(|\mathbb{M}(n)^{2}| \textrm{ } ; \textrm{ } k,m,n\Big) - \mathbb{E}\Big(|\mathbb{M}(n)| \textrm{ } ; \textrm{ } k,m,n\Big)^{2}.$$

It is important to note that when $m=1$ the variance is identically $0$. Therefore we will calculate the variance for $m \geq 2$.

Recall from Corollary \ref{fd_sec_mom_bounds} that 

$$\ineql{n}{2k}{\mathbb{C}}{\mathbb{L}}{2} \leq \mathbb{E}\Big(|\mathbb{M}(n)^{2}| \textrm{ } ; \textrm{ } k,m,n\Big) \leq \inequ{n}{2k}{\mathbb{C}}{\mathbb{U}}{2}$$ 

and from Corollary \ref{fd_variance}

$$\ineql{n}{2k}{\mathbb{C}^{2}}{\mathbb{L}}{1} \leq \Big(\mathbb{E}\Big(|\mathbb{M}(n)| \textrm{ } ; \textrm{ } k,m,n\Big)\Big)^{2} \leq \inequ{n}{2k}{\mathbb{C}^{2}}{2\mathbb{U}}{1}$$ 

by applying Lemma \ref{lemma6} (occurs in Section \ref{chapter:formulae}).

Thus the variance is bounded above by

\begin{align*}
\mathbb{VAR} \Big( |\mathbb{M}(n)| \textrm{ } ; \textrm{ } k,m,n )\Big ) &\leq  n^{2k}\mathbb{C}_{2} + n^{2k-1}\mathbb{C}_{2}\mathbb{U}_{2} - n^{2k}\mathbb{C}_{1}^{2} + n^{2k-1}\mathbb{C}_{1}^{2}\mathbb{L}_{1}\\ \\
	&\leq  n^{2k} (\mathbb{C}_{2} -\mathbb{C}_{1}^{2}) + n^{2k-1}(\mathbb{C}_{2}\mathbb{U}_{2} + \mathbb{C}_{1}^{2}\mathbb{L}_{1})\\ \\
	&\leq n^{2k} (\mathbb{C}_{2} -\mathbb{C}_{1}^{2})  \Big( 1 + \frac{\mathbb{C}_{2}\mathbb{U}_{2} - \mathbb{C}_{1}^{2}\mathbb{L}_{1}}{n(\mathbb{C}_{2} -\mathbb{C}_{1}^{2})}\Big).
\end{align*}
and bounded below by 
\begin{align*}
\mathbb{VAR} \Big(|\mathbb{M}(n)| \textrm{ } ; \textrm{ } k,m,n )\Big ) &\geq  n^{2k}\mathbb{C}_{2} - n^{2k-1}\mathbb{C}_{2}\mathbb{L}_{2} - n^{2k}\mathbb{C}_{1}^{2} - n^{2k-1}\mathbb{C}_{1}^{2}\mathbb{U}_{1}\\ \\
	& \geq  n^{2k} (\mathbb{C}_{2} -\mathbb{C}_{1}^{2}) - n^{2k-1}(\mathbb{C}_{2}\mathbb{L}_{2} - \mathbb{C}_{1}^{2}\mathbb{U}_{1})\\ \\
	& \geq n^{2k} (\mathbb{C}_{2} -\mathbb{C}_{1}^{2})  \Big( 1 + \frac{\mathbb{C}_{2}\mathbb{L}_{2}}{n(\mathbb{C}_{2} -\mathbb{C}_{1}^{2})}\Big).
\end{align*}
since $\mathbb{L}_{1} = 0$. Recall that
$$\mathbb{C}_{1}^{2} = \Big(\frac{1}{(m+k-1)_{k}}\Big)^{2}$$ and $$\mathbb{C}_{2} = \frac{2^{k}}{(m+2k-1)_{2k}}.$$ Then $$\frac{2^{k}}{(m+2k-1)_{2k}} - \frac{1}{((m+k-1)_{k})^{2}} > 0$$ since $$ 2^{k}((m+k-1)_{k})^{2} > (m+2k-1)_{2k}.$$
\end{proof}

\begin{corollary}\label{bound_exp}
Let $k \geq 1$ be fixed, $n \geq  m \geq k^{2}$, then the expected number of instances of subnetwork motifs of size $k$ in Full Duplication is $$(\frac{n}{m})^{k} \Big(1-\frac{k^{2}}{2m}\Big) \leq \mathbb{E}\Big(|\mathbb{M}(n)| \textrm{ } ; \textrm{ } k,m,n\Big) \leq (\frac{n}{m})^{k}\Big(1+\frac{3k^{2}}{4n}\Big).$$
\end{corollary}

\begin{proof}

From the result of Corollary \ref{fulld_var_bounds} we know that

$$\frac{n^{k}}{(m+k-1)_{k}} \leq \mathbb{E}\Big(|\mathbb{M}(n)| \textrm{ } ; \textrm{ } k,m,n\Big) \leq \frac{n^{k}}{(m+k-1)_{k}}(1 + \frac{3k^{2}}{4n})$$
when $m$ is fixed. 
We can apply Lemma \ref{lemma4} (occurs in Section \ref{chapter:formulae}) to approximate the bounds for \\ {$1/(m+k-1)_{k}$} when $m$ is growing. Doing so we obtain $$\frac{1}{m^{k}} (1-\frac{k^{2}}{2m}) \leq \frac{1}{(m+k-1)_{k}} \leq \frac{1}{m^{k}}.$$
\end{proof}

\begin{corollary}\label{corollary6}
Let $k \geq 1$ be fixed, $n \geq  m \geq k^{2}$, then the second moment of the number of instances of subnetwork motifs of size $k$ in Full Duplication is

$$2^{k}(\frac{n}{m})^{2k}(1-\frac{4m^{2}}{n})(1 - \frac{2k^{2}}{m}) \leq \mathbb{E}\Big(|\mathbb{M}(n)^{2}| \textrm{ } ; \textrm{ } k,m,n \Big) \leq 2^{k}(\frac{n}{m})^{2k}(1+\frac{9m^{2}}{n})$$
\end{corollary}

\begin{proof}
We know from Corollary \ref{fd_sec_mom_bounds} that the second moment is
$$\frac{n^{2k}2^{k}}{(m+2k-1)_{2k}}(1-\frac{4m^{2}}{n}) \leq \mathbb{E} \Big(|\mathbb{M}(n)^{2}| \textrm{ } ; \textrm{ } k,m,m \Big) \leq \frac{n^{2k}2^{k}}{(m+2k-1)_{2k}}(1+\frac{9m^{2}}{n})$$
\noindent when $m$ is fixed. 
Using Lemma \ref{lemma4} (occurs in Section \ref{chapter:formulae}) we can bound $1/(m+2k-1)_{2k}$ as m grows. Doing so we obtain

$$\frac{1}{m^{2k}} (1 - \frac{2k^{2}}{m}) \leq \frac{1}{(m+2k-1)_{2k}} \leq \frac{1}{m^{2k}} .$$
\end{proof}

\begin{corollary}
Let $k \geq 1$ be fixed, $n \geq  m \geq k^{2}$, then the variance of the number of instances of subnetwork motifs of size $k$ in Full Duplication is

$$\Big(\frac{n}{m}\Big)^{2k}\Big( 2^{k} - 1\Big)\Big( 1 - \frac{2^{k+3}+3}{(2^{k}-1)m} - \frac{m^{2}}{(2^{k}-1)n}\Big) \leq \mathbb{VAR}\Big(|\mathbb{M}(n)| \textrm{ } ; \textrm{ } k,m,n) \Big)$$

$$\mathbb{VAR}\Big(|\mathbb{M}(n)| \textrm{ } ; \textrm{ } k,m,n) \Big) \leq \Big(\frac{n}{m}\Big)^{2k}\Big( 2^{k} - 1\Big)\Big( 1 +\frac{k^{2}}{(2^{k}-1)m}+ \frac{2^{k}9m^{2}}{(2^{k} -1)n} \Big).$$
\end{corollary}

\begin{proof}
In order to calculate the variance we will use the standard expression of variance as stated below
$$\mathbb{VAR} \Big(|\mathbb{M}(n)| \textrm{ } ; \textrm{ } k,m,n \Big) = \mathbb{E}\Big(|\mathbb{M}(n)^{2}| \textrm{ } ; \textrm{ } k,m,n\Big) - \mathbb{E}\Big(|\mathbb{M}(n)| \textrm{ } ; \textrm{ } k,m,n\Big)^{2}.$$

Using the result of Corollary \ref{bound_exp_fixedm} we will calculate the bounds of the square of the first moment 

$$\Big(\frac{n}{m}\Big)^{2k}\Big(1-\frac{k^{2}}{m}\Big) \leq \mathbb{E}\Big(|\mathbb{M}(n)| \textrm{ } ; \textrm{ } k,m,n\Big)^{2} \leq \Big(\frac{n}{m}\Big)^{2k}\Big(1+\frac{3k^{2}}{2n}\Big)$$

where the lower bound error follows from Lemma \ref{lemma4} (occurs in Section \ref{chapter:formulae}) and the upper bound error follows from Lemma \ref{lemma10} (occurs in Section \ref{chapter:formulae}). 

Using the result of Corollary \ref{corollary6} (occurs in Section \ref{chapter:formulae}) we can calculate bounds for the variance as follows,

\begin{align*}\mathbb{VAR} \Big(|\mathbb{M}(n)| \textrm{ } ; \textrm{ } k,m,n \Big) &\leq 2^{k}\Big(\frac{n}{m}\Big)^{2k}\Big(1+\frac{9m^{2}}{n}\Big) - \Big(\frac{n}{m}\Big)^{2k}\Big(1-\frac{k^{2}}{m}\Big) \\ \\
&\leq \Big(\frac{n}{m}\Big)^{2k} \Big( 2^{k} - 1 + \frac{k^{2}}{m} + \frac{9m^{2}2^{k}n^{2k}}{n}\Big) \\ \\ 
&\leq  \Big(\frac{n}{m}\Big)^{2k}\Big( 2^{k} - 1\Big)\Big( 1 +\frac{k^{2}}{(2^{k}-1)m}+ \frac{2^{k}9m^{2}}{(2^{k} -1)n} \Big),
\end{align*}

\begin{align*}\mathbb{VAR} \Big(|\mathbb{M}(n)| \textrm{ } ; \textrm{ } k,m,n \Big) &\geq 2^{k}\Big(\frac{n}{m}\Big)^{2k}\Big(1-\frac{4m^{2}}{n}\Big)\Big(1-\frac{2k^{2}}{m}\Big) - \Big(\frac{n}{m}\Big)^{2k}\Big(1+\frac{3k^{2}}{2n}\Big) \\ \\
&\geq 2^{k} \Big(\frac{n}{m}\Big)^{2k} \Big(1 - \frac{2k^{2}}{m} - \frac{4m^{2}}{n}\Big) - \Big(\frac{n}{m}\Big)^{2k}\Big(1+\frac{3k^{2}}{2n}\Big) \\ \\
&\geq \Big(\frac{n}{m}\Big)^{k}\Big(2^{k} - 1\Big)\Big(1 - \frac{2^{k}2k^{2}}{m(2^{k} -1)} - \frac{2^{k}8m^{2} + 3k^{2}}{2n(2^{k} -1)}\Big) \\ \\ 
&\geq \Big(\frac{n}{m}\Big)^{k}\Big(2^{k} - 1\Big)\Big( 1 - \frac{2^{k+3}+3}{(2^{k}-1)m} - \frac{m^{2}}{(2^{k} -1)n}\Big).
\end{align*}
\end{proof}

\section{Partial Duplication}
The Partial Duplication mode occurs under the $\it{random \textrm{ } duplication \textrm{ } and \textrm{ } inheritance \textrm{ } process}$, which is an extension of the gene duplication process that involves random inheritance at every step that is controlled by a vector of probabilities, $\vec{\pi}$. At the end of the random duplication and inheritance process there is a set of subnetwork motifs given $m,n,k,\vec{\pi}$ and this section will cover the expectation of that. 
 
A $\it{Partial \textrm{ } Duplication}$ mode is an inheritance model that covers the expectation at some stage $n$ of a subnetwork motif $\mathbb{M}$ for some arbitrary k. The significance of this section is that the expected number of instances of $\mathbb{M}$ depends on a vector of inheritance probabilities $\vec{0} \leq \vec{\pi} \leq \vec{1}$. Recall that each $\mathbb{M}$ has an associated vector of probabilities $\vec{\pi} = (\pi_{1}, ..., \pi_{k})$. However, the previous section covers a special case of this inheritance mode where $\vec{\pi} = \vec{1}$. It'll turn out that $\vec{\pi}$ completely determines the expected number of subnetwork motifs for any given stage $n$ in duplication given the random duplication and inheritance process.

\subsection{First Moments}
To begin evaluating the expected number of subnetwork motif instances of any size $k$, we will look at instances of single gene subnetwork motifs where $k=1$. Let $s$ be the size of the family at any stage $n$ and $p$ be the probability of inheritance. In this case, we define $f(p,s)$ as the expectation of a single gene subnetwork motif $\mathbb{M}$. The expected number of instances of these single gene subnetwork motifs only depends on $p$ and $s$ which follows from the proof of the following lemma. We will show that the expectation can be expressed as a ratio of gamma functions and we will later show that we can express the expectation as a generating function which will be useful when we let $k \geq 1$.

\begin{lemma}
\label{lemma_self_loops_PD}
Assume $\mathbb{M}$ is a single gene subnetwork motif. Let $s$ be the size of the gene family and $p$ be the probability of inheritance. Then the expected number of instances of $\mathbb{M}$ under Partial Duplication is 
$$f(p,s) = \frac{\Gamma(p+s)}{\Gamma(s)\Gamma(p+1)}.$$
\end{lemma}

\begin{proof}
We will prove the statement of the lemma by induction on $s$.

Assume there have been no duplications in the family belonging to $\mathbb{M}$; then the only possible single gene subnetwork motif is the original instance of the subnetwork motif. If $s=1$ then 
$$f(p,1) = \frac{\Gamma(p+1)}{\Gamma(1)\Gamma(p+1)} = 1$$
since $\Gamma(1) = 1$.

Assume the induction hypothesis is true for some $s \geq 1$. We know that when the family size is $s$ the expected number of instances are $f(p,s)$ and the probability of adding one gene to the family belonging to $\mathbb{M}$ upon duplication is $\frac{f(p,s)}{s} \cdot p$ since each gene has an equal chance of being selected for duplication. Thus, 
$$f(p,s+1) = f(p,s) + f(p,s)\cdot \frac{p}{s} = f(p,s)\Big(1 + \frac{p}{s}\Big) = \frac{p+s}{s}f(p,s)$$

Using the induction hypothesis the expectation is as follows

$$\frac{p+s}{s}f(p,s) =\frac{p+s}{s} \cdot \frac{\Gamma(p+s)}{\Gamma(s)\Gamma(p+1)} = \frac{\Gamma(p+s+1)}{\Gamma(s+1)\Gamma(p+1)}.$$
\end{proof}

Suppose $\mathbb{M}$ is a subnetwork motif of arbitrary size $k \geq 1$. Assume as in section 3.1 that for notational convenience the families that belong to $\mathbb{M}$ are indexed from $1 \textrm{ to } k$. To begin evaluating the expected number of subnetwork motif instances of size $k \geq 1$, let $\vec{s} = (s_{1}, ..., s_{k})$ where $s_{i}$ is the size of the $i^{th}$ family at any stage $n$ for $i =1, ..., k$ and $\vec{\pi} = (\pi_{1}, ..., \pi_{k})$ gives the inheritance probabilities associated with $\mathbb{M}$. 

\begin{theorem}
\label{partial_duplication_motif_exp_corollary}
Suppose $\mathbb{M}$ is a subnetwork motif of arbitrary size $k \geq 1$. Let $k \leq m \leq n$ and $\vec{0} \leq \vec{\pi} \leq \vec{1}$ be fixed. Let $\vec{s} =(s_{1},...,s_{k})$ be the sequence of family sizes belong to $\mathbb{M}$ at some stage $n$. Then the expected number of instances of subnetwork motifs conditioned on the sequence of family sizes is

$$\mathbb{E}\Big(|\mathbb{M}(\vec{s})| \Big) = \prod_{i=1}^{k}f(\pi_{i},s_{i}).$$

\end{theorem}

\begin{proof}
In order to prove the theorem we will be inducting on $h = ||\textrm{ }\vec{s}\textrm{ }||$. Assume that $h=k$. Then $\vec{s} = (1, ..., 1)$, so the only possible instance is the original instance and $|\mathbb{M}(\vec{1})| = 1$. Since we know from the proof of Lemma $\ref{lemma_self_loops_PD}$ that $f(\pi_{i}, 1) =1$ for $i=1, ...,k$ we see that $$\prod_{i=1}^{k}f(\pi_{i}, 1)  = 1.$$ Thus the base case is verified.

Assume that the theorem holds true for some $h \geq k$. We will now show that it holds true for $h+1$. Let $\vec{s^{\prime}} =  (s_{1}, ...., s_{j-1}, s_{j}^{\prime}, s_{j+1}, ..., s_{k})$ be a sequence of family sizes such that $||\vec{s^{\prime}}|| = h + 1$. Thus at least one duplication has occurred. Without loss of generality assume that the most recent duplication occurred in the $j^{th}$ family. Just prior to the last duplication $\vec{s} = (s_{1}, ...., s_{j-1}, s_{j}, s_{j+1}, ..., s_{k})$ is the sequence of family sizes such that $||\vec{s}|| = h$, $s_{j} = s_{j}^{\prime} -1$, and $\mathbb{E}(|\mathbb{M}(\vec{s})|) = \prod_{i=1}^{k} f(\pi_{i}, s_{i})$ by the induction hypothesis. When the gene is duplicated from the $j^{th}$ family  each instance of $\mathbb{M}(\vec{s})$ has probability $\frac{\pi_{j}}{s_{j}}$ of giving rise to a new instance therefore the expected number of new instances is $\Big(\prod_{i=1}^{k} f(\pi_{i}, s_{i})\Big)\Big(\frac{\pi_{j}}{s_{j}}\Big)$. Given the recursion we proved in Lemma \ref{lemma_self_loops_PD} and $s_{j}^{\prime} = s_{j}+1$  the total expectation is as follows, 

\begin{align*}
\mathbb{E}\Big(|\mathbb{M}(\vec{s^{\prime}})| \Big) &= \mathbb{E}\Big(|\mathbb{M}(\vec{s})|\Big)+ \mathbb{E}\Big(|\mathbb{M}(\vec{s})|\Big) \cdot \frac{\pi_{j}}{s_{j}} \\
&= \mathbb{E}\Big(|\mathbb{M}(\vec{s})|\Big)\Big(1 + \frac{\pi_{j}}{s_{j}}\Big) \\
&= \Big(\prod_{i=1}^{k} f(\pi_{i}, s_{i})\Big)\Big(1 +\frac{\pi_{j}}{s_{j}}\Big) \\
&=  \prod_{i=1}^{k} f(\pi_{i}, s_{i}^{\prime})
\end{align*}

since $s_{i} = s_{i}^{\prime}$ if $i \neq j$. 

\end{proof}

We will now evaluate what happens to the expectation of the number of instances of $\mathbb{M}(n)$ when the set is not conditioned on the vector of family sizes. For the purposes of this calculation it is useful to begin with the following Corollary. 

\begin{corollary}
Assume $\mathbb{M}$ is a single gene subnetwork motif. Let $s$ be the size of the gene family and $p$ be the probability of inheritance. Then the expected number of instances of $\mathbb{M}$ under Partial Duplication can be expressed as

$$f(p,s) = [x^{s}]\Big(\frac{x}{(1-x)^{p+1}}\Big)$$
\end{corollary}

\begin{proof}
We know from Lemma \ref{lemma_self_loops_PD}
 that 
$$f(p,s) = \frac{\Gamma(p+s)}{\Gamma(s)\Gamma(p+1)}$$
Now we can apply Lemma 1 we obtain the generating function, 
\begin{align*}
f(p,s) &= \frac{\Gamma(p+s)}{\Gamma(s)\Gamma(p+1)} \\ \\ &= \frac{(p+s-1)_{s-1}}{(s-1)!} \\ \\ &= [x^{s-1}]\Big(\frac{1}{(1-x)^{p+1}}\Big) \\ \\ &= [x^{s}]\Big(\frac{x}{(1-x)^{p+1}}\Big).
\end{align*}
\end{proof}

\begin{theorem}\label{pdexpthm}
Suppose $\mathbb{M}$ is a subnetwork motif of size $k$, $1 \leq k \leq m \leq n$, $\vec{0} \leq \vec{\pi} \leq \vec{1}$ is fixed, $\textrm{and } \hat{\pi} = \pi_{1} + ... + \pi_{k}$. Then the expected number of instances of $\mathbb{M}$ in Partial Duplication is 
$$\mathbb{E}\Big(|\mathbb{M}(n)\textrm{ }|\textrm{ }m,n,\vec{\pi},k\Big) = \frac{\Gamma(\hat{\pi} + n)\Gamma(m)}{\Gamma(\hat{\pi}+m)\Gamma(n)}.$$
\end{theorem}
This result allows us to construct a significance test for subnetwork motifs using the mean of the number of instances of $\mathbb{M}$ under Partial Duplication. 

\begin{proof}
In order to prove the statement of the theorem we will take a similar approach from Theorem \ref{full_dup_fixed_n_1st_mom} and utilize generating functions. We know that 
$$\mathbb{E}\Big(\prod_{i=1}^{k} f(\pi_{i},s_{i})\Big) = \binom{n-1}{m-1}^{-1}\Big( \sum_{h=k}^{n-m+k} \binom{n-h-1}{m-k-1}\sum_{h} \prod_{i=1}^{k}f(\pi_{i},s_{i})\Big),$$

where $s_{1}, ..., s_{k}$ are the family sizes, $h = s_{1} + ... + s_{k}$, and $\vec{s}$ is a composition of $h$ into $k$ parts.

From Theorem 3 we know the generating function associated with $\sum_{h=k}^{n-m+k} \binom{n-h-1}{m-k-1}$ is 

$$a(x) = (x+x^{2}+ ...)^{m-k} = \Big(\frac{x}{1-x}\Big)^{m-k}.$$

The generating function associated with $\sum_{h} \prod_{i=1}^{k}f(\pi_{i},s_{i})$ is 

\begin{align}\label{PDgenfun}
\prod_{i=1}^{k}\Big(\frac{x}{(1-x)^{\pi_{i} +1}}\Big).
\end{align}	

We will now utilize the generating function to obtain the expectation by dividing the coeffecient of $x^{n}$ in Expression $\eqref{PDgenfun}$ by the total number of equally likely compositions of $n$ into $m$ parts. Thus, 

\begin{align*}
\mathbb{E}\Big(\prod_{i=1}^{k} f(\pi_{i},s_{i})\Big) &= \Big( [x^{n}] \big( \frac{x^{m}}{(1-x)^{\hat{\pi} +m}}\big)\Big) \binom{n-1}{m-1}^{-1} \\ \\ 
&= \Big( [x^{n-m}]\big(\frac{1}{(1-x)^{\hat{\pi} +m}}\big)\Big)  \binom{n-1}{m-1}^{-1} \\ \\
&= \frac{\binom{\hat{\pi} +n -1}{n-m}}{\binom{n-1}{m-1}} \\ \\
&= \frac{\Gamma(\hat{\pi} + n)\Gamma(m)}{\Gamma(\hat{\pi}+m)\Gamma(n)}.
\end{align*}
\end{proof}

In order to derive bounds for the expected number of subnetwork motifs in Partial Duplication the following lemmas will be useful.

\begin{lemma}\label{pdexplemma}
Suppose $z\geq 0, y \geq 1$ and $y \geq 2z^{2}$ then 
$$z \ln(y) - \frac{1}{y}\big(\frac{z}{2} + \frac{1}{12}\big) \leq \ln \Big(\frac{\Gamma(y+z)}{\Gamma(y)}\Big) \leq z \ln(y) + \frac{z^{2}}{y}$$
\end{lemma}

\begin{proof}
In order to derive bounds for the ratio we will use the fact that
$$\ln(\Gamma(x))  = (x - \frac{1}{2}) \ln(x) - x + \frac{1}{2}\ln(2\pi) + \sum_{a=1}^{\infty} \frac{\mathrm{B}_{2a}}{2a(2a-1)x^{2a-1}}$$ for $x > 0$ \citep{mathfun}, where $B_{2a}$ are Bernoulli numbers and $B_{2} = \frac{1}{6}$. The above expression can be bounded by truncating the summation at the first term neglected such that when the first term neglected is negative an upper bound is obtained and when the first term neglected is positive a lower bound is obtained. Then 

$$(x-\frac{1}{2})\ln(x) - x + \frac{1}{2}\ln(2\pi) \leq \ln(\Gamma(x)) \leq (x-\frac{1}{2})\ln(x) - x + \frac{1}{2}\ln(2\pi) + \frac{1}{12x}.$$

To obtain an upper bound for $\ln\Big(\frac{\Gamma(y+z)}{\Gamma(y)}\Big)$ we derive an upper bound for $\ln(\Gamma(y+z))$, a lower bound for $\ln(\Gamma(y))$, and then combine them. We can apply the asymptotic relationship when $x = y+z$ and when $x=y$.
The upper bound calculation is as follows:
\begin{align*}
 \ln \Big(\frac{\Gamma(y+z)}{\Gamma(y)}\Big) &= \ln(\Gamma(y + z)) - \ln(\Gamma(y)) \\
 &\leq (z+y - \frac{1}{2})\ln(z+y) - z - y + \frac{1}{2}\ln(2\pi) + \frac{1}{12(y+z)} - (y - \frac{1}{2})\ln(y) + y - \frac{1}{2}\ln(2\pi) \\
 &= (z+y - \frac{1}{2})\ln(z+y) - z + \frac{1}{12(y+z)} - (y - \frac{1}{2})\ln(y) \\ 
 &\leq z \ln(z+y) + (y - \frac{1}{2})\big(\ln(z+y) - \ln(y)\big) - z -  \frac{1}{12(y+z)} \\
 &=  z \ln(z+y) + (y - \frac{1}{2})\big(\ln(1 + \frac{z}{y})\big) - z - \frac{1}{12(y+z)} \\
 &=  z \ln(z+y) + (y - \frac{1}{2})\big(\frac{z}{y} - \frac{z^{2}}{2y} + \frac{z^{3}}{3y} ... \big) - z - \frac{1}{12(y+z)}\\
 &\leq z \ln(z+y) + (y - \frac{1}{2})\big(\frac{z}{y}\big) - z - \frac{1}{12(y+z)}\\
 &=  z \ln(y(\frac{z}{y} + 1))  + (y - \frac{1}{2})\big(\frac{z}{y}\big) - z -  \frac{1}{12(y+z)}\\
&\leq z \ln(y) + \frac{z^{2}}{y} - \frac{z}{2y} +  \frac{1}{12(y+z)} \\
&\leq z \ln(y) + \frac{z^{2}}{y}.
\end{align*}

To derive a lower bound we derive a lower bound for $\ln(\Gamma(y+z))$, an upper bound for $\ln(\Gamma(y))$, and then combine them. Then the lower bound is as follows: 
\begin{align*}
 \ln \Big(\frac{\Gamma(y+z)}{\Gamma(y)}\Big) &= \ln(\Gamma(y + z)) - \ln(\Gamma(y)) \\
 &\leq (z+y - \frac{1}{2})\ln(z+y) - (y - \frac{1}{2})\ln(y) - z - \frac{1}{12y} \\
 &= z\ln(y+z) + (y-\frac{1}{2})\ln(z+y) - (n-\frac{1}{2})\ln(n) - z - \frac{1}{12y} \\
 &= z\ln(y+z) + (y-\frac{1}{2})\big(\ln(1+\frac{z}{y})\big) - z - \frac{1}{12y} \\
 &\leq z\ln(y+z) + (y-\frac{1}{2})\big(\frac{z}{n} - \frac{z^{2}}{2y^{2}}\big) - z - \frac{1}{12y} \\
 &=  z\ln(y) + z(\frac{z}{y} - \frac{z^{2}}{2y^{2}}) + (y-\frac{1}{2})\big(\frac{z}{y} - \frac{z^{2}}{2y^{2}}\big) - z - \frac{1}{12y} \\
 &= z\ln(y) + \frac{1}{n}\big( z^{2} - \frac{z^{3}}{2n} - \frac{z^{2}}{2} - \frac{z}{2} + \frac{z^{2}}{4} - \frac{1}{12}\big) \\
 &\geq z \ln(y) - \frac{1}{y}\big(\frac{3z^{2}}{4} + \frac{z^{3}}{2n} + \frac{z}{2} + \frac{1}{12}\big) \\
 &\geq z\ln(y) - \frac{1}{y}\big(\frac{z}{2} + \frac{1}{12}\big).
\end{align*}
\end{proof}

\begin{lemma}\label{pdexplemma_expo}
Suppose $z \geq 0, y \geq 1,$ and $y \geq 2z^{2}$ then

$$y^{z}(1 - \frac{1}{y}(\frac{z}{2} + \frac{1}{12})) \leq \frac{\Gamma(y+z)}{\Gamma(y)}  \leq y^{z}(1+\frac{3z^{2}}{y}).$$
\end{lemma}

\begin{proof}
In order to obtain the result of the Lemma we begin by exponentiating the result of Lemma \ref{pdexplemma} to obtain

$$y^{z}e^{-(1/y)(z/2 + 1/12)}  \leq \frac{\Gamma(y+z)}{\Gamma(y)} \leq y^{z}e^{z^{2}/y}.$$

By applying Lemma \ref{explemma} (occurs in Section \ref{chapter:formulae}) with $\zeta = \frac{z^{2}}{y}$ to the upper bound and Lemma \ref{negexplemma} with $\zeta = \frac{1}{y}(\frac{z}{2} + \frac{1}{12})$ to the lower bound we obtain the result of the Lemma. Note that the conditions for Lemma \ref{explemma} and \ref{negexplemma} (both occur in Section \ref{chapter:formulae}) follow directly from the hypothesis. 
\end{proof}

\begin{corollary}\label{pdexbounds}
Let $k\geq 1$ be fixed, $n \geq m \geq 1, n \geq 2k^{2}$, and $0 \leq ||\textrm{ }\vec{\pi}\textrm{ }|| \leq k$. Then the expected number of instances of $\mathbb{M}(n)$ in Partial Duplication is 

$$\frac{n^{||\vec{\pi}||}\Gamma(m)}{\Gamma(|| \vec{\pi} || + m)}\Big(1-\frac{1}{n}\big(\frac{k}{2} + \frac{1}{12}\big)\Big) \leq \frac{\Gamma(|| \vec{\pi} || +n) \Gamma(m)}{\Gamma(n)\Gamma(|| \vec{\pi} || +m)} \leq  \frac{n^{||\vec{\pi}||}\Gamma(m)}{\Gamma(||\vec{\pi}|| + m)} \Big(1 + \frac{3k^{2}}{2n} \Big).$$
\end{corollary}

\begin{proof}
We know from Theorem \ref{pdexpthm} that $\mathbb{E}\Big(|\mathbb{M}(n)|\textrm{ }m,n,\vec{\pi},k\Big)$ is 
$$\frac{\Gamma(||\vec{\pi}|| + n)\Gamma(m)}{\Gamma(||\vec{\pi}||+m)\Gamma(n)}.$$

We derive the indicated bounds for the expectation by applying Lemma \ref{pdexplemma_expo} with $y = n$ and $z = ||\vec{\pi}||$. 

Note that we obtain the following more informative bounds by applying Lemma \ref{pdexplemma_expo} to $\mathbb{E}\Big(|\mathbb{M}(n)|\textrm{ }m,n,\vec{\pi},k\Big)$ when $0 \leq ||\vec{\pi}|| < \frac{1}{2}$

$$\frac{n^{||\vec{\pi}||}\Gamma(m)}{\Gamma(||\vec{\pi}|| + m)}\Big(1 - \frac{31(||\vec{\pi}||)^{2}}{32n}\Big) \leq \frac{\Gamma(||\vec{\pi}|| +n) \Gamma(m)}{\Gamma(n)\Gamma(||\vec{\pi}|| +m)} \leq  \frac{n^{||\vec{\pi}||}\Gamma(m)}{\Gamma(||\vec{\pi}|| + m)} \Big(1 - \frac{(||\vec{\pi}||^{2}}{n} + \frac{||\vec{\pi}||^{4}}{2n^{2}} \Big).$$
\end{proof}

\begin{lemma}\label{recip_pdexplemma}
Suppose $z \geq 0, y \geq 1,$ and $y \geq 2z^{2}$ then
$$-z \ln(y) + \frac{1}{y}(\frac{z}{2} + \frac{1}{12}) \leq \ln \big(\frac{\Gamma(y)}{\Gamma(y+z)}\big) \leq -z \ln(y) - \frac{z^{2}}{y}.$$
\end{lemma}

\begin{proof}
The result of the Lemma follows directly from multiplying the result of Lemma \ref{pdexplemma} by $-1.$
\end{proof}

\begin{lemma}\label{recip_pdexplemma_expo}
Suppose $z \geq 0, y \geq 1,$ and $y \geq 2z^{2}$ then
$$y^{-z}(1+\frac{1}{y}(\frac{z}{2} + \frac{1}{12})) \leq \frac{\Gamma(y)}{\Gamma(y+z)} \leq y^{-z}(1-\frac{z^{2}}{y} + \frac{z^{4}}{2y^{2}}).$$
\end{lemma}

\begin{proof}
To obtain the result of the Lemma we begin by exponentiating the result of Lemma \ref{recip_pdexplemma} to obtain 

$$y^{-z} e^{(1/y)(z/2 +1/12)} \leq \frac{\Gamma(y)}{\Gamma(y+z)} \leq y^{-z}e^{-z^{2}/y}.$$

We obtain the result of the Lemma by applying Lemma \ref{negexplemma} (occurs in Section \ref{chapter:formulae}) with $\zeta = \frac{z^{2}}{y}$ to the upper bound and Lemma \ref{explemma} (occurs in Section \ref{chapter:formulae})  with $\zeta = \frac{1}{y}(\frac{z}{2} + \frac{1}{12})$ to the lower bound. 
\end{proof}

\begin{corollary}
Let $k\geq 1$ be fixed, $n \geq m \geq 1, n \geq 2k^{2}, \frac{m}{n^{2}} \to \infty$, and $0 \leq ||\vec{\pi}|| \leq k$. Then the expected number of instances of $\mathbb{M}(n)$ in Partial Duplication is 

$$\frac{\Gamma(||\vec{\pi}|| +n) \Gamma(m)}{\Gamma(n)\Gamma(||\vec{\pi}|| +m)} \geq \Big(\frac{n}{m}\Big)^{||\vec{\pi}||} \Big(1-\frac{1}{n}\big(\frac{k}{2} + \frac{1}{12}\big)\Big)\Big( 1 + \frac{1}{m}(\frac{||\vec{\pi}||^{2}}{2} + \frac{1}{12})\Big)$$

$$\frac{\Gamma(||\vec{\pi}|| +n) \Gamma(m)}{\Gamma(n)\Gamma(||\vec{\pi}|| +m)} \leq \Big(\frac{n}{m}\Big)^{||\vec{\pi}||} \Big(1 + \frac{3k^{2}}{2n} \Big)\Big( 1 - \frac{||\vec{\pi}||^{2}}{m} + \frac{||\vec{\pi}||^{4}}{2m^{2}}\Big).$$
\end{corollary}

\begin{proof}
The indicated bounds follow directly from applying Lemma \ref{recip_pdexplemma_expo} with \\$y=m$ and $z=||\vec{\pi}||$ to the result of Corollary \ref{pdexbounds}.
\end{proof}

\subsection{Second Moments, $\lowercase{k}=1$}
Now we turn our attention to studying the second moments of the number of instances of a given subnetwork motif. We will begin the study by considering the case of single gene subnetwork motifs. In order to calculate the variance we will need to analyze the expected number of ordered pairs of single gene subnetwork motifs. 

Let $\mathbb{M}$ be an arbitrary single gene subnetwork motif with inheritance probability $p = \pi_{1}$ and family size $s.$ For the purposes of this subsection, we will be conditioning on $s$ so we let $\mathbb{M}(s)$ be the set of instances of $\mathbb{M}$ when the family size is $s.$ The members of the of the subnetwork motif's gene family are indexed in order of duplication and denoted  $\{b_{1}, ..., b_{s}\}.$ Here, $(b_{1})$ is the original instance of $\mathbb{M}$ and $\{(b_{2}), ..., (b_{s})\}$ are the potential members of $\mathbb{M}(s)$.

The expected size of $\mathbb{M}(s)$ is $f(p,s)$ by Lemma 10. Let $g(p,s)$ be the expected size of $\mathbb{M}(s)^{2}$. We now prove a recurrence satisfied by $g(p,s).$ 

\begin{corollary}\label{solo_gps_recur}
We have $g(p,1) = 1$ and for all $s\geq1$
$$g(p, s+1) = g(p,s) + \frac{2p}{s}\cdot g(p,s) + \frac{p}{s}\cdot f(p,s).$$
\end{corollary}

\begin{proof}
When the family size is $s=1$ no duplications have occurred. At that point the only pair of subnetwork motif instances is the original instance of $\mathbb{M}$ and itself, so $g(p,1)=1$. 

We will now show that the recurrence relationship is true for any $s \geq 1$. Any member $<(b_{i}),(b_{j})>  \in \mathbb{M}(s+1)^{2}$ belongs to one of three categories: the pair falls into the first category if $1 \leq i,j \leq s$, the second category if $1 \leq i \leq s\textrm{ and }j=s+1\textrm{ or }i=s+1\textrm{ and }1 \leq j \leq s$, and the third category if $ i = j = s+1$. In order to calculate $g(p,s+1)$ we will find the expected number of ordered pairs of each category. 

In the first category the pair consists of two instances of $\mathbb{M}$ when the family size is $s$. Thus the expected number of such pairs is $g(p,s)$ by definition. 

In the second category the pair consists of an old subnetwork motif instance and a new subnetwork motif instance. A pair from the second category can only be generated from a pair from the first category by duplicating one the of elements of the pair. Suppose $<(b_{i}),(b_{j})> \in \mathbb{M}(s)^{2}$. To generate a pair in the second category $b_{i}$ can be duplicated and the subnetwork motif must be inherited in duplication. There is a $\frac{1}{s}$ chance that $b_{i}$ is selected for duplication to generate $b_{s+1}$ and probability $p$ the new subnetwork motif instance $(b_{s+1})$ is inherited in duplication. We sum over all the pairs in $\mathbb{M}(s)^{2}$ to calculate the expected number of category 2 pairs in $\mathbb{M}(s+1)^{2}$. Since the expected number of pairs in $\mathbb{M}(s)^{2}$ is $g(p,s)$ the expected number of pairs in category 2 where the first element is the new instance is $\frac{p}{s}\cdot g(p,s)$. Note that the expected number of pairs in category 2 where the second element is the new instance is also $\frac{p}{s}\cdot g(p,s)$. Since the events are disjoint, $\frac{2p}{s} \cdot g(p,s)$ is the expected number of category 2 pairs in $\mathbb{M}(s+1)^{2}$. 

In the third category the only possible pair is the new instance and itself. The new instance of $\mathbb{M}$ must be generated from a member of $\mathbb{M}(s)$. Suppose $(b_{i}) \in \mathbb{M}(s)$. Then there is a $\frac{1}{s}$ chance that $b_{i}$ is selected for duplication to generate the new gene $b_{s+1}$ and the new gene has probability $p$ of becoming a new instance $(b_{s+1})$. To find the expected number of category 3 pairs we sum over all the members in $\mathbb{M}(s)$. By Lemma \ref{lemma_self_loops_PD} the expected number of members in $\mathbb{M}(s)$ is $f(p,s)$ so the expected number of category 3 pairs in $\mathbb{M}(s+1)^{2}$ is $\frac{p}{s}\cdot f(p,s).$

The corollary follows from the fact that the sum of expectations is the expectation of the sum.
\end{proof}

\begin{corollary}\label{g_gen_fun}
For $s \geq 1$ we have
$$g(p,s) = \big( \frac{2\Gamma(s + 2p)}{\Gamma(s)\Gamma(2p+1)} - \frac{\Gamma(p+s)}{\Gamma(s)\Gamma(p+1)}\big).$$
\end{corollary}

\begin{proof}
Let $h(p,s) = g(p,s) + f(p,s)$. Then $h(p,1) = 2$. For $s \geq 1$ we have the following:
\begin{align*}
   h(p,s+1) &= g(p,s+1) + f(p,s+1) \\ 
   &= g(p,s) + \frac{2p}{s}g(p,s) + \frac{p}{s}f(p,s) + f(p,s)(1+\frac{p}{s}) \\
   &= h(p,s) + \frac{2p}{s}g(p,s) + \frac{2p}{s}f(p,s) \\
   &= h(p,s) + \frac{2p}{s}h(p,s) \\
   &= h(p,s)\big(1 + \frac{2p}{s}\big).
\end{align*}

Therefore by induction on $s$
$$h(p,s) = 2 \prod_{i=1}^{s-1} \big(1 +\frac{2p}{i}\big) = \frac{2 \Gamma(s+2p)}{\Gamma(s)\Gamma(2p+1)} \text{ for all } s \geq 1.$$

Thus $$g(p,s) = \big( \frac{2\Gamma(s + 2p)}{\Gamma(s)\Gamma(2p+1)} - \frac{\Gamma(p+s)}{\Gamma(s)\Gamma(p+1)}\big)$$ since we know $f(p,s)$ from Lemma \ref{lemma_self_loops_PD} and $g(p,s) = h(p,s) - f(p,s)$. 
\end{proof}

\begin{corollary}\label{gps_gen_fun}
For $s \geq 1$ we have
$$g(p,s) = [x^{s}]\Big( \frac{2x}{(1-x)^{2p+1}} - \frac{x}{(1-x)^{p+1}}\Big).$$
\end{corollary}
\begin{proof}
We know from the proof of Corollary  \ref{g_gen_fun} that $h(p,s) = \frac{2\Gamma(s + 2p)}{\Gamma(s)\Gamma(2p+1)}.$ Thus 
\begin{align*}
h(p,s) &= \frac{2(s+2p -1)_{s-1}}{(s-1)} \\ \\
&= \binom{2(s+2p -1)}{s-1} \\ \\
&= [x^{s-1}]\big( \frac{2}{(1-x)^{2p+1}} \big) \\ \\
&= [x^{s}] \big( \frac{2x}{(1-x)^{2p+1}}\big),
\end{align*}
using Lemma 1. Then
$$g(p,s) = [x^{s}]\Big( \frac{2}{(1-x)^{2p+s}}\Big) - [x^{s}] \Big( \frac{x}{(1-x)^{p+1}}\Big)$$

\noindent from Lemma \ref{lemma_self_loops_PD} and the fact that $g(p,s) = h(p,s) - f(p,s)$.
\end{proof}

Notice that the value of $g(p,s)$ depends only on the values of $p \text{ and } s$. That is to say the expected number of pairs of subnetwork motifs of size $k=1$ depends only on $p$ and $s$. In the next section we will show by example that when $k \geq 2$ the size of the gene families and corresponding inheritance probabilities do not necessarily determine the expected number of pairs of subnetwork motifs.

\subsection{Maximizing the Second Moment}

We will now study the second moment when $k \geq 2$. In order to do so we must specify how subnetwork motifs are inherited when they share a common gene which is duplicated. Assume as in previous sections that for notational convenience the families that belong to $\mathbb{M}$ are indexed from $1$ to $k$. The following inheritance modes are two of the possible refinements of the basic Partial Duplication mode. 

Previously we aggregated our random duplication process over the $k$ family sizes that contribute to $\mathbb{M}$. Now we need a finer analysis and we will aggregate according to the sequence of duplications. It will be useful to select an arbitrary sequence of duplications for the purposes of computation. At the start of the duplication process there are $m$ genes to choose from. After the first duplication there are $m+1$ genes to choose from, and the options for genes increment after every duplication. Given $n-m$ duplications there are $(m)(m+1)(m+2) \cdot ... \cdot (m+n-1) = (n-1)_{n-m}$ possible sequence of duplications, and they are all equally likely since the duplication at each stage is chosen at random.

The genes are indexed as follows. For $j=1, ..., m$, $u_{j}$ is the original gene in the $j^{th}$ family. For $m <j \leq n$, $u_{j}$ is the gene that results from the $(j-m)^{th}$ duplication. For $m \leq j \leq n$ let $U_{j} = \{  u_{\ell} \textrm{ } |\textrm{ } 1 \leq \ell \leq j\}.$ Then $U_{j}$ is the set of genes which exist after the $(j-m)^{th}$ duplication. For $1 \leq i \leq k$ let $S_{i}$ be the set of genes in $U_{n}$ which belong to the $i^{th}$ family which contributes to $\mathbb{M}$. Then $\mathbb{M}(n) \subseteq S_{1} \times ... \times S_{k}.$ That is to say that the set of potential subnetwork motifs is a subset of the product of the $S_{i}$'s.

Suppose $\mathbb{M}$ is a subnetwork motif of arbitrary size $k \geq 2$, $\mathscr{I}$ and $\mathscr{J}$ are instances of $\mathbb{M}(n)$ that share a particular gene $b$ from the $i^{th}$ family for $1 \leq i \leq k$, and $b$ is duplicated to generate $b^{\prime}$. In the first refinement when $b^{\prime}$ is generated the inheritance events $\mathscr{I} \in \mathbb{M}(n)$ and $\mathscr{J} \in \mathbb{M}(n)$ are fully correlated, so that $\pi_{i}$ is the probability that both inheritance events occur. In the second refinement when when $b^{\prime}$ is generated the probability that the inheritance events $\mathscr{I}^{\prime} \in \mathbb{M}(n)$ and $\mathscr{J}^{\prime} \in \mathbb{M}(n)$ occur are mutually independent, so that $\pi_{i}^{2}$ is the probability that both inheritance events occur.

\begin{figure}
\centering
    \includegraphics[height=18cm, width=\textwidth]{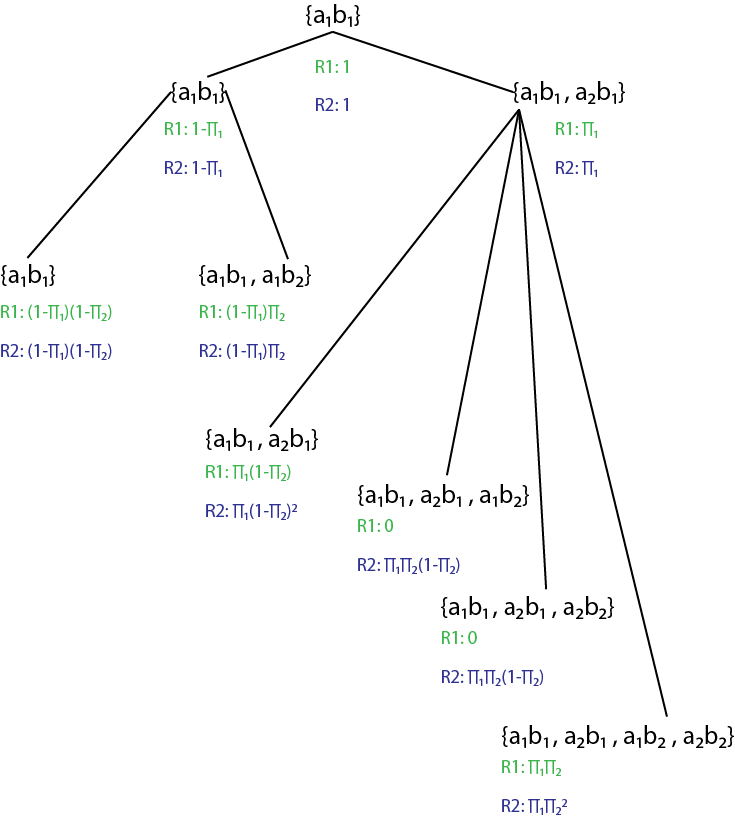}
    \caption[Outcomes for Refinement Comparison]{The duplication process begins at stage 0 with the original subnetwork motif $\{a_{1},b_{1}\}$. If the sequence of genes after the duplication process is $\{a_{1}, b_{1}, a_{2}, b_{2}\}$ then all possible subnetwork motif events are depicted with probabilities for each. The probabilities in green (on top) are from refinement 1 and the probabilities in blue (on the bottom) are from refinement 2.}
    \label{fig:first}
\end{figure}

It can be seen in Figure \ref{fig:first} that given a particular sequence of duplications, these two refinements give different results for the second moment. Take the simple case of $\vec{s}=(2,2)$ and $\vec{\pi} = (\frac{1}{2}, \frac{1}{2})$. The second moment in the first refinement is $6 \frac{1}{4}$ and the second moment in the second refinement is $5\frac{3}{4}$. 

We choose to focus on the second refinement, defined below as $\it{Binary\textrm{ }Inheritance}$, because it is tractable and gives us the maximum second moment value of any refinement of the Partial Duplication mode. The latter will be proved in Theorem \ref{maxBI}.

\begin{definition}
Binary inheritance is a refinement of the Partial Duplication mode which acts as follows. Suppose $\mathbb{M}$ is a subnetwork motif of size $k, 1 \leq k \leq m \leq n, \mathscr{I}\textrm{ and }\mathscr{J}$ are instances of $\mathbb{M}$ that share a common gene $b$ in the $i^{th}$ family for $1\leq i \leq k$, and $\vec{0} \leq \vec{\pi} \leq \vec{1}$ is fixed. For any arbitrary duplication step, if $b$ is selected for duplication to generate $b^{\prime}$ then with probability $\pi_{i}$ the inheritance events $\mathscr{I}^{\prime} \in \mathbb{M}(n)$ and $\mathscr{J}^{\prime} \in \mathbb{M}(n)$ both occur, and with probability $1 - \pi_{i}$ neither of the inheritance events occurs. 
\end{definition}

Note that if the common gene $b$ is contained in $r$ instances then all $r$ possible inheritance events occur or none of them occur. Also if $i >k$ there are no possible inheritance events that can occur at that step. 
\begin{theorem}\label{maxBI}
Suppose $\mathbb{M}$ is a subnetwork motif of size $k, 1 \leq k \leq m \leq n$, and $\vec{0} \leq \vec{\pi} \leq \vec{1}$ is fixed. Then Binary Inheritance gives the maximum value for $\mathbb{E}\Big(|\mathbb{M}(n)^{2}| ; m,n,\vec{\pi}, k \Big)$ over all refinements of Partial Duplication.
\end{theorem}

This result allows us to obtain the variance of the number of instances of $\mathbb{M}$ under Partial Duplication using the Binary Inheritance mode. 

\begin{proof}
Let Challenge Inheritance be an arbitrary refinement of Partial Duplication which we denote by $\mathbb{C}.$ Then let $\mathbb{E}(|\mathbb{M}(n)^{2}|\textrm{ } ;\textrm{ } \mathbb{C})$ be the expected number of pairs of subnetwork motif instances under Challenge Inheritance for the given $m, n, \vec{\pi}, \textrm{ and } k$. Let Binary Inheritance be denoted by $\mathbb{B}$ and let $\mathbb{E}(|\mathbb{M}(n)^{2}|\textrm{ };\textrm{ }\mathbb{B})$ be the expected number of pairs of subnetwork motif instances under Binary Inheritance. We will show that 

\begin{align}\label{BgeqCsecmom}
\mathbb{E}(|\mathbb{M}(n)^{2}|\textrm{ };\textrm{ } \mathbb{C}, m,n, \vec{\pi}_{i}, k) \leq \mathbb{E}(|\mathbb{M}(n)^{2}|\textrm{ } ;\textrm{ } \mathbb{B}, m,n, \vec{\pi}_{i}, k).
\end{align}

In order to prove the statement of the theorem we fix on one sequence of duplications and calculate the expected number of ordered pairs  in $\mathbb{M}(n)$ for refinements $\mathbb{C}$ and $\mathbb{B}$. We are going to consider the probability that pairs of potential instances are contained in $\mathbb{M}(n)$ conditioned on the chosen sequence of $n-m$ duplications. 

We choose an arbitrary ordered pair of potential subnetwork motif instances $(\mathscr{I}, \mathscr{H})$, $\emph{i.e.,}$ $\mathscr{I}, \mathscr{H} \in S_{1} \times ... \times S_{k}.$ Suppose $\mathscr{I} \in \mathbb{M}(n)$. For $m \leq j \leq n$ there is a predecessor to $\mathscr{I}$ which we call $\mathscr{I}_{j}$, which must belong to $\mathbb{M}(j)$ in order for $\mathscr{I}$ to be in $\mathbb{M}(n)$. We will define this sequence by inducting on $j = n$ down to $j=m$.

To begin suppose $\mathscr{I}_{n} = \mathscr{I}\textrm{ then }\mathscr{I} _{n} \in \mathbb{M}(n).$ For $m\leq j < n\textrm{ let }\hat{u}$ be the gene from which $u_{j+1}$ is duplicated and assume that $\mathscr{I}_{j+1} \in \mathbb{M}(j+1)$. In the case that $u_{j+1}$ does not appear in $\mathscr{I}_{j+1}$, let $\mathscr{I}_{j} = \mathscr{I}_{j+1}$; thus it can be seen that $\mathscr{I}_{j+1} \in \mathbb{M}(j)$. In the case that $u_{j+1}$ does appear in $\mathscr{I}_{j+1}\textrm{, let }\mathscr{I}_{j}$ be an instance that is obtained by replacing $u_{j+1}$ with $\hat{u}$ and $\mathscr{I}_{j} \in \mathbb{M}(j)$.

From the inductive definition of the sequence of subnetwork motif instances it can be seen that the members of $\mathscr{I}_{j}$ are contained in $U_{j}$ for $m \leq j \leq n$. In particular the members of $\mathscr{I}_{m}$ must be contained in $U_{m}$; thus it consists only of original genes and must be the original instance of $\mathbb{M}$. 

It is easy to see by induction on $j=n$ to $j=m$ that the definition and results for $(\mathscr{I}_{m}, ..., \mathscr{I}_{n})$ apply equally to $(\mathscr{H}_{m}, ..., \mathscr{H}_{n})$. We can now express the probability that $\mathscr{I}$ and $\mathscr{H}$ are both in $\mathbb{M}(n)$ as 

\begin{align}
Pr\big( \mathscr{I}_{m}, \mathscr{H}_{m} \in \mathbb{M}(m) \big)\prod_{j=m}^{n-1} Pr\big(\mathscr{I}_{j+1}, \mathscr{H}_{j+1} \in \mathbb{M}(j+1) \textrm{ } |\textrm{ } \mathscr{I}_{j}, \mathscr{H}_{j} \in \mathbb{M}(j)\big).
\end{align}

Since $\mathscr{I}_{m}$ and $\mathscr{H}_{m}$ are the original instance of $\mathbb{M}$ the probability that $\mathscr{I}_{m}$ and $\mathscr{H}_{m}$ are contained in $\mathbb{M}(n)$ is 1. For any $m \leq j \leq n-1$, we are interested in the conditional probability that $\mathscr{I}_{j+1}, \mathscr{H}_{j+1} \in \mathbb{M}(j+1)$ given $\mathscr{I}_{j}, \mathscr{H}_{j} \in \mathbb{M}(j)$. So assume that $\mathscr{I}_{j} \textrm{ and } \mathscr{H}_{j}$ are contained in $\mathbb{M}(j)$. At stage $j$, let $\hat{u}$ be selected for duplication to produce the new gene $u_{j+1}$ in the $i^{th}$ family. If $u_{j+1}$ does not appear in $\mathscr{I}_{j+1}$ nor $\mathscr{H}_{j+1}$ then the probability that both $\mathscr{I}_{j+1}, \mathscr{H}_{j+1} \in \mathbb{M}(j+1)$ is 1. If $u_{j+1}$ is in only in $\mathscr{I}_{j+1}$ or $\mathscr{H}_{j+1}$ then the probability that both $\mathscr{I}_{j+1}, \mathscr{H}_{j+1} \in \mathbb{M}(j+1)$ is $\pi_{i}$. Note that if a duplication occurs outside of the first $k$ families then $\mathbb{M}(j+1) = \mathbb{M}(j)$ and the probability is 1. If $u_{j+1}$ appears in both $\mathscr{I}_{j+1}$ and $\mathscr{H}_{j+1}$ then the probability that both $\mathscr{I}_{j+1}, \mathscr{H}_{j+1} \in \mathbb{M}(j+1)$ under $\mathbb{C}$ is between $0$ and $\pi_{i}$, whereas under $\mathbb{B}$ the probability is exactly $\pi_{i}$. Note that the only time the probability that $\mathscr{I}_{j+1}, \mathscr{H}_{j+1} \in \mathbb{M}(j+1)$ can differ between $\mathbb{C}$ and $\mathbb{B}$ is in this case. It follows that 

\begin{align}\label{BgeqCpair}
Pr\big(\mathscr{I}, \mathscr{H} \in \mathbb{M}(n) \textrm{ } ; \textrm{ } \mathbb{C} \big) \leq Pr\big(\mathscr{I}, \mathscr{H} \in \mathbb{M}(n) \textrm{ } ; \textrm{ } \mathbb{B} \big).
\end{align}
Summing Equation \ref{BgeqCpair} over $\mathscr{I}, \mathscr{H} \in S_{1} \times ... \times S_{k}$ we obtain
$$\sum_{\mathscr{I}, \mathscr{H}} Pr\big(\mathscr{I}, \mathscr{H} \in \mathbb{M}(n)\textrm{ ; } \mathbb{C}\big) \leq \sum_{\mathscr{I}, \mathscr{H}} Pr\big(\mathscr{I}, \mathscr{H} \in \mathbb{M}(n)\textrm{ ; } \mathbb{B}\big)$$

\noindent so that

$$\mathbb{E}(|\mathbb{M}(n)^{2}|\textrm{ };\textrm{ } \mathbb{C}) \leq \mathbb{E}(|\mathbb{M}(n)^{2}|\textrm{ } ;\textrm{ } \mathbb{B}).$$

\noindent Recall that there are $(n-1)_{n-m}$ possible sequences of duplications. The results above are conditioned on the arbitrarily chosen sequence of duplications. In order to extend the results to the overall second moment we average overall all $(n-1)_{n-m}$ sequences to obtain Equation \eqref{BgeqCsecmom}, which is the desired result.

From now on we will confine our study of Partial Duplication to the Binary Inheritance refinement.
\end{proof}

\subsection{Second Moments for Binary Inheritance}
We will now study second moments for Binary Inheritance. For the purposes of this section we will adopt the gene indexing and terminology from the introduction to Theorem \ref{maxBI}. Thus we will be aggregating our gene duplication process over the sequence of duplications rather than the family sizes that contribute to $\mathbb{M}$. In addition, we will consider properties of particular realizations of the duplication and inheritance process that includes not only the selection of what to duplicate but also which particular instances are inherited under Binary Inheritance. 

Suppose $\mathbb{M}$ is a subnetwork motif of size $1 \leq k \leq m \leq n$ and $\vec{\pi} = (\pi_{1}, ..., \pi_{k})$ is the vector of inheritance probabilities. As before, $\vec{S} = (S_{1}, ..., S_{k})$ is the sequence of the sets of genes that belong to $\mathbb{M}$ where $S_{i} \subseteq U_{n}$ consist of the genes in the $i^{th}$ family. We define a gene $\hat{u} \in U_{n}$ to be $\it{viable}$ if and only if $\hat{u}$ appears in a member of $\mathbb{M}(n)$. If $\hat{u} = u_{j}$ for $1 \leq j \leq k$, then $\hat{u}$ is viable since it appears in the original instance of $\mathbb{M}$. Alternatively, if $\hat{u} = u_{j}$ for $m \leq j \leq n$ then $\hat{u}$ will be viable if and only if $u_{j}$ appears in a member of $\mathbb{M}(j)$. When $j \geq m$ a new instance of $\mathbb{M}$ can only be inherited from an instance of $\mathbb{M}$ and a viable gene can only be inherited from a viable gene. Note that if a gene is viable at the stage when it appears then it remains viable throughout the duplication process. 

\begin{lemma}\label{viability}
A k-tuple of $S_{1} \times ... \times S_{k}$ belongs to  $\mathbb{M}(n)$ if and only if every gene in it is viable. \end{lemma}

\begin{proof}
Every instance in $\mathbb{M}(n)$ consists of viable genes by the definition of viability. 

It remains to prove that if we have a $k$-tuple of viable genes $\mathscr{L} \in S_{1} \times ... \times S_{k}$ then it forms an instance of $\mathbb{M}$. Let $j$ be the smallest value such that $j \geq m$ and all the members of $\mathscr{L}$ are in $U_{j}$. The proof proceeds by induction on $j$. 

Suppose $j=m$. Then there have been no duplications. Thus the only possibility for $\mathscr{L}$ is the original instance of $\mathbb{M}$.
Suppose $j > m$ and let $\mathscr{L} = (a_{1}, ...., a_{i-1}, a_{i}, a_{i+1}, ..., a_{k})$. By the minimality of $j$ we know that $u_{j} = a_{i}$ for some $ i \in \mathbb{K}$ where $\mathbb{K} = \{1, ..., k\}$. In order for $a_{i}$ to be viable it must be inherited from a viable gene which we denote $\hat{a}_{i}$. Since $\hat{a}_{i}$ is viable then $\mathscr{I} = (a_{1}, ..., a_{i-1}, \hat{a_{i}}, a_{i+1}, ..., a_{k})$ must be an instance of $\mathbb{M}$ and every member of $\mathscr{I}$ must belong to $U_{j-1}$.

By the induction hypothesis $\mathscr{I}$ is an instance of $\mathbb{M}$. A viable gene can only be inherited from a viable gene, therefore $a_{i}$ must appear in some member of  $\mathbb{M}$.  Since $a_{i}$ appears in an instance of $\mathbb{M}$ it must have been inherited from $\hat{a}_{i}$ which appears in another instance of $\mathbb{M}$. Under the Binary Inheritance Mode, when $\hat{a}_{i}$ is selected for duplication to generate $a_{i}$, either every instance that contains $\hat{a}_{i}$ results in a new instance or there are no new instances. Therefore, it follows that $\mathscr{L}$ must be inherited from $\mathscr{I}$. 
\end{proof}

\begin{corollary}\label{sameprob}
Suppose $\mathbb{M}$ is a subnetwork motif of size $k \geq 1$ under the Binary Inheritance Mode, $(S_{1}, ..., S_{k})$ is the sequence of the gene families belonging to $\mathbb{M}$, $s_{i} = |S_{i}|$ for $i \in \mathbb{K}$, and $\vec{\pi} = (\pi_{1}, ..., \pi_{k})$ is the vector of inheritance probabilities. Then in family $S_{i}$ $f(\pi_{i}, s_{i})$ is the expected number of viable genes and $g(\pi_{i}, s_{i})$ is the expected number of ordered pairs of viable genes.
\end{corollary}

\begin{proof}
Within $S_{i}$ a viable gene can only be inherited from a viable gene and has chance $\pi_{i}$ of inheriting viability. Therefore, viability is inherited by the same probabilistic process as single gene subnetwork motifs and we can apply Lemma \ref{lemma_self_loops_PD} and Corollary \ref{g_gen_fun} to obtain the expected number of viable genes and the second moment for viable genes respectively. 
\end{proof}

\begin{corollary}\label{ex_g_pi_s}
Assume the same hypothesis as Corollary \ref{sameprob}. Then the mean number of ordered pairs of instances of $\mathbb{M}$ conditioned on the vector of family sizes is
$$\mathbb{E}\big(|\mathbb{M}(n)|^{2} \textrm{ ; }s_{1}, ..., s_{k} \big) = \prod_{i=1}^{k}g(\pi_{i}, s_{i}).$$
\end{corollary}

\begin{proof}
Let $V_{i}$ be the set of viable genes in $S_{i}$ for $1 \leq i \leq k$. By Lemma \ref{viability} we know that $\mathbb{M}(n) = V_{1} \times ... \times V_{k}$. Then $\mathbb{E}\big(|\mathbb{M}(n)|^{2} \textrm { ; } s_{1},...,s_{k}\big) = \mathbb{E}\big(\prod_{i=1}^{k} |V_{i}^{2}|\big)$.
We know by Corollary \ref{sameprob} that the expectation of $|V_{i}^{2}|$ is $g(\pi_{i}, s_{i})$ for $i=1, ...,k$. The expected number of ordered pairs of $\mathbb{M}$ is 
$$\mathbb{E}\big(|\mathbb{M}(n)|^{2}\textrm{ ; }s_{1}, ..., s_{k}\big) = \mathbb{E}\big(\prod_{i=1}^{k} |V_{i}|^{2}\big) = \prod_{i=1}^{k} \mathbb{E}\big(|V_{i}|^{2}\big) = \prod_{i=1}^{k} g(\pi_{i}, s_{i}),$$
since the probability of viability for any gene is independent from the probability of viability for any gene in a different family. 
\end{proof}

\begin{theorem}\label{BIsecmom}
Suppose $\mathbb{M}$ is a subnetwork motif of size $k$, $1\leq k \leq m \leq n$ and $\text{ }\vec{0} \leq \vec{\pi} \leq \vec{1}$ is fixed. Then the second moment of the number of instances of $\mathbb{M}$ in binary Partial Duplication is

$$\mathbb{E}\Big(|\mathbb{M}(n)|^{2} ; m,n,\vec{\pi},k\Big) = \frac{\Gamma(m)}{\Gamma(n)} \sum_{\mathbb{A} \subseteq \mathbb{K}} \frac{(-1)^{|\mathbb{A}|} 2^{k}\Gamma(n+ ||\vec{\pi}|| + \sum_{i \notin \mathbb{A}} \pi_{i})}{2^{|\mathbb{A}|}\Gamma(m+||\vec{\pi}|| + \sum_{i \notin \mathbb{A}}\pi_{i})}.$$
\end{theorem}
This result allows us to calculate the variance of the number of instances of $\mathbb{M}$ under Partial Duplication using Binary Inheritance, which would be required for a significance test. 
\begin{proof}
Our approach to prove the statement of the Theorem is similar to our approach to proving Theorem \ref{full_dup_fixed_n_1st_mom} by utilizing generating functions to calculate the value of $\mathbb{E}\Big(|\mathbb{M}(n)|^{2} ; m,n,\vec{\pi},k\Big)$. Corollary \ref{sameprob} gives the following expression for the second moment

$$\mathbb{E}\Big(|\mathbb{M}(n)|^{2} ; m,n,\vec{\pi},k\Big) = \binom{n-1}{m-1}^{-1} \Big( \sum_{h=k}^{n-m+k} \binom{n-h-1}{m-k-1} \sum_{||\vec{s}||} \prod_{i=1}^{k} g(\pi_{i}, s_{i})\Big),$$

\noindent where the inner summation is over $||\vec{s}|| = h$. From the proof of Lemma \ref{exp_lemma} we know that 

$$[x^{h}]\sum_{h=k}^{n-m+k} \binom{n-h-1}{m-k-1} = (x + x^{2} + ...)^{m-k} = (\frac{x}{(1-x)})^{m-k}.$$

From Corollaries \ref{gps_gen_fun} and \ref{ex_g_pi_s} we obtain

\begin{align*}
\sum_{h \geq 1} x^{h} \sum \prod_{i=1}^{k} g(\pi_{i}, s_{i}) &= \prod_{i=1}^{k} \Big( \frac{2x}{(1-x)^{2\pi_{i} + 1}} - \frac{x}{(1-x)^{\pi_{i} + 1}}\Big) \\
&= \sum_{\mathbb{A} \subseteq \{ \mathbb{K} \} }\Big( \prod_{i \in \mathbb{K} - \mathbb{A}}  \frac{2x}{(1-x)^{2\pi_{i} + 1}}\Big) \Big( \prod_{j \in \mathbb{A}} - \frac{x}{(1-x)^{\pi_{j} + 1}}\Big)
\end{align*}

\noindent the second sum is over $||\vec{s}|| = h$ and $\mathbb{K} = \{1, ..., k\}$. 

As in Theorem \ref{full_dup_fixed_n_1st_mom} we can apply the generating function to obtain the expectation by dividing the coefficient of $x^{n}$ in the product of the generating functions by the total number of equally likely compositions of $n$ into $m$ parts. That is, \\ $\mathbb{E}\Big(|\mathbb{M}(n)|^{2} ; m,n,\vec{\pi},k\Big) \times \binom{n-1}{m-1}^{-1}$ is 

\begin{align*}
& [x^{n}]\Big( (\frac{x}{(1-x)})^{m-k}\sum_{Q \subseteq  \mathbb{K}  }\Big( \prod_{i \in \mathbb{K} - \mathbb{A}}  \frac{2x}{(1-x)^{2\pi_{i} + 1}}\Big) \Big( \prod_{j \in  \mathbb{A}} - \frac{x}{(1-x)^{\pi_{j} + 1}}\Big)\Big) \\ \\
&=[x^{n}] \sum_{\mathbb{A} \subseteq \mathbb{K} } \frac{x^{m-k} (2x)^{|\mathbb{K} - \mathbb{A}|} (-x)^{|\mathbb{A}|}} {(1-x)^{m-k}(1-x)^{2(\sum_{i \notin \mathbb{A}}\pi_{i}) + |\mathbb{K} - \mathbb{A}|}(1-x)^{\sum_{j \in \mathbb{A}}\pi_{j} + | \mathbb{A}|}} \\ \\
&=[x^{n-m}] \sum_{\mathbb{A} \subseteq  \mathbb{K}} \frac{(-1)^{\mathbb{A}} 2^{k}} {(1-x)^{m+2(\sum_{i \notin \mathbb{A}} \pi_{i}) + \sum_{j \in \mathbb{A}} \pi_{j}}}\\ \\ 
&= \sum_{\mathbb{A} \subseteq \mathbb{K}} \frac{(-1)^{|\mathbb{A}|} 2^{k}\Gamma(n+ ||\vec{\pi}|| + \sum_{i \notin \mathbb{A}} \pi_{i})}{2^{|\mathbb{A}|}\Gamma(m+||\vec{\pi}|| + \sum_{i \notin \mathbb{A}}\pi_{i})}.
\end{align*}

We obtain the result of the theorem by solving for the expectation and simplifying.
\end{proof}

\begin{corollary}
Suppose $\mathbb{M}$ is a subnetwork motif of size $k$, $1\leq k \leq m \leq n$ and $\vec{0} \leq \vec{\pi} \leq \vec{1}$ is fixed. Then the variance of the number of instances of $\mathbb{M}$ in binary Partial Duplication is $\mathbb{VAR}\Big(|\mathbb{M}(n)| ; m,n,\vec{\pi},k\Big)$ which evaluates to
$$\frac{\Gamma(m)}{\Gamma(n)} \sum_{Q \subseteq \mathbb{K}} \frac{(-1)^{k - |Q|} 2^{|Q|}\Gamma(n+ ||\vec{\pi}|| + \sum_{i \in Q} \pi_{i})}{\Gamma(m+||\vec{\pi}|| + \sum_{i \in Q}\pi_{i})}- \Big(\frac{\Gamma(||\vec{\pi}|| + n)\Gamma(m)}{\Gamma(||\vec{\pi}|| + m)\Gamma(n)}\Big)^{2}.$$
\end{corollary}

This result allows us to construct a significance test for $\mathbb{M}$ under Partial Duplication using Binary Inheritance.

\begin{proof}
The Corollary follows immediately from Theorems \ref{pdexpthm} and \ref{BIsecmom}. 
\end{proof}

\begin{lemma}\label{bonlemma}
Suppose $a_{i} > b_{i} \geq 0$ for $1 \leq i \leq k$, and for $0 \leq j \leq k$ let 

$$\mathbb{S}_{j} = \sum_{\mathbb{A} \subseteq \mathbb{K}} \Big( \prod_{i \in \mathbb{K} - \mathbb{A}} a_{i} \prod_{i \in \mathbb{A}} b_{i}\Big)$$

where the summation is over $|\mathbb{A}_{i}| = j$. Then 

$$\mathbb{S}_{0} - \mathbb{S}_{1} + \mathbb{S}_{2} - ... - \mathbb{S}_{2\ell+1} \leq \prod_{i \in \mathbb{K}} (a_{i} - b_{i}) \leq \mathbb{S}_{0} - \mathbb{S}_{1} + ... + \mathbb{S}_{2\ell},$$

for  $0 \leq \ell \leq \frac{k}{2}.$
\end{lemma}

\begin{proof}
Since 
$$\mathbb{S}_{0} = \prod_{i \in \mathbb{K}} a_{i} > 0$$ 
the following inequalities, 
$$1 - \frac{\mathbb{S}_{1}}{\mathbb{S}_{0}} + \frac{\mathbb{S}_{2}}{\mathbb{S}_{0}} - ... - \frac{\mathbb{S}_{2\ell+1}}{\mathbb{S}_{0}} \leq \prod_{i \in \mathbb{K}} (1 - \frac{b_{i}}{a_{i}}) \leq 1 - \frac{\mathbb{S}_{1}}{\mathbb{S}_{0}} + ... + \frac{\mathbb{S}_{2\ell}}{\mathbb{S}_{0}},$$

\noindent are equivalent to the inequalities in the statement of the the lemma and are implied by the Bonferroni inequalities \citep{feller}. To see the latter, simply apply the Bonferroni inequalities to $k$ mutually independent events $E_{1}, ..., E_{k}$ where $E_{i}$ has probability $\frac{b_{i}}{a_{i}}$ for $i=1, ..., k$.
\end{proof}

\begin{corollary}\label{BIbounds}
Suppose $\mathbb{M}$ is a subnetwork motif of size $k$, $1 \leq k \leq m \leq n,$ and $\vec{0} \leq \vec{\pi} \leq \vec{1}$. Then the second moment of the number of instances of subnetwork motifs of size $k$ in partial binary duplication satisfies 

$$\mathbb{E}\Big(|\mathbb{M}(n)|^{2} ; m,n,\vec{\pi},k\Big) \geq \frac{\Gamma(m)2^{k}\Gamma(n + 2 ||\vec{\pi}||)}{\Gamma(n)\Gamma(m + 2 ||\vec{\pi}||)} - \frac{\Gamma(m)}{\Gamma(n)} \sum_{j=1}^{k} \frac{2^{k-1} \Gamma(n + 2 ||\vec{\pi}|| - \pi_{j})}{ \Gamma(m + 2 ||\vec{\pi}|| - \pi_{j})}$$
and 
$$\mathbb{E}\Big(|\mathbb{M}(n)|^{2} ; m,n,\vec{\pi},k\Big) \leq \frac{\Gamma(m)2^{k}\Gamma(n + 2 ||\vec{\pi}||)}{\Gamma(n)\Gamma(m + 2 ||\vec{\pi}||)}.$$

\end{corollary}

\begin{proof}
Our approach to prove the statement of the Corollary is to find bounds for the second moment given a particular $\vec{s}$ and then use them to bound the result of Theorem \ref{BIsecmom}. 

Recall from the proof of Theorem \ref{BIsecmom} that
$$\mathbb{E}\Big(|\mathbb{M}(n)|^{2} ; m,n,\vec{\pi},k\Big) = \binom{n-1}{m-1}^{-1} \Big( \sum_{h=k}^{n-m+k} \binom{n-h-1}{m-k-1} \sum \prod_{i=1}^{k} g(\pi_{i}, s_{i})\Big)$$
where the second sum is over $||\vec{s}|| = h$.

To obtain bounds for $\mathbb{E}\Big(|\mathbb{M}(n)|^{2} ; m,n,\vec{\pi},k\Big)$ we start by applying Lemma \ref{bonlemma} to $ \prod_{i=1}^{k} g(\pi_{i}, s_{i})$ where $a_{i} = h(\pi_{i}, s_{i})$ and $b_{i} = f(\pi_{i}, s_{i})$ since \\ $g(\pi_{i}, s_{i}) = h(\pi_{i}, s_{i}) - f(\pi_{i}, s_{i})$. Thus the simplest bounds for $\prod_{i=1}^{k} g(\pi_{i}, s_{i})$ are, 

$$\mathbb{S}_{0} - \mathbb{S}_{1} \leq \prod_{i=1}^{k} g(\pi_{i}, s_{i}) \leq \mathbb{S}_{0}$$

\noindent where $\mathbb{S}_{0} = \prod_{i=1}^{k} h(\pi_{i}, s_{i})$ and $\mathbb{S}_{1} =\sum_{j=1}^{k} f(\pi_{j}, s_{j})\prod_{i=1,i \neq j}^{k} h(\pi_{i}, s_{i}).$

By replacing $\binom{n-h-1}{m-k-1}$ with its generating function and using the bounds for $\prod_{i=1}^{k} g(\pi_{i}, s_{i})$ we see
$$\mathbb{E}\Big(|\mathbb{M}(n)|^{2} ; m,n,\vec{\pi},k\Big) > \binom{n-1}{m-1}^{-1} [x^{n}]\Big( (\frac{x}{(1-x)})^{m-k} \big(\mathbb{S}_{0} - \mathbb{S}_{1}\big)\Big)$$
and
$$\mathbb{E}\Big(|\mathbb{M}(n)|^{2} ; m,n,\vec{\pi},k\Big) < \binom{n-1}{m-1}^{-1} [x^{n}]\Big( (\frac{x}{(1-x)})^{m-k} \mathbb{S}_{0}\Big).$$

Thus we can use the generating functions for $h(\pi_{i}, s_{i})$ and $f(\pi_{j}, s_{j})$ to calculate upper and lower bounds for the second moment. The upper bound is as follows
\begin{align*}
\mathbb{E}\Big(|\mathbb{M}(n)|^{2} ; m,n,\vec{\pi},k\Big) &<  \binom{n-1}{m-1}^{-1}  [x^{n}]\Big( (\frac{x}{(1-x)})^{m-k}\big(  \prod_{i=1}^{k} h(\pi_{i}, s_{i}\big)\Big) \\
&=  \binom{n-1}{m-1}^{-1}  [x^{n}]\Big( (\frac{x}{(1-x)})^{m-k}\Big( \prod_{i=1}^{k}  \frac{2x}{(1-x)^{2\pi_{i} + 1}}\Big)\Big) \\
&=\binom{n-1}{m-1}^{-1}  [x^{n}]\Big( (\frac{x}{(1-x)})^{m-k}\Big( \prod_{i=1}^{k}  \frac{2x}{(1-x)^{2\pi_{i} + 1}}\Big) \\
&= \binom{n-1}{m-1}^{-1} [x^{n-m}]\frac{2^{k}}{(1-x)^{m + 2||\vec{\pi}||}} \\
&=\frac{\Gamma(m)2^{k}\Gamma(n + 2 ||\vec{\pi}||)}{\Gamma(n)\Gamma(m + 2 ||\vec{\pi}||)}.
\end{align*}
To obtain results for the lower bound we calculate the second term in the lower bound since the first term is the upper bound. Thus,
\begin{align*}
& \binom{n-1}{m-1}^{-1}  [x^{n}]\Big( (\frac{x}{(1-x)})^{m-k}\Big( \sum_{j=1}^{k} f(\pi_{j}, s_{j})\prod_{i=1,i \neq j}^{k} h(\pi_{i}, s_{i})\Big) \\
&= \binom{n-1}{m-1}^{-1}  [x^{n}]\Big( (\frac{x}{(1-x)})^{m-k}\Big( \sum_{j=1}^{k} \frac{x}{(1-x)^{\pi_{i} +1}}\prod_{i=1, 1 \neq j}^{k}  \frac{2x}{(1-x)^{2\pi_{i} + 1}}\Big) \\
&= \binom{n-1}{m-1}^{-1} \sum_{j=1}^{k} [x^{n-m}] \frac{2^{k-1}}{(1-x)^{m -1 + (2 ||\vec{\pi}|| - \pi_{j})}} \\
&= \frac{\Gamma(m)}{\Gamma(n)} \sum_{j=1}^{k} \frac{2^{k-1} \Gamma(n + 2 ||\vec{\pi}|| - \pi_{j})}{ \Gamma(m + 2 ||\vec{\pi}|| - \pi_{j})}.
\end{align*}

Therefore the lower bound is as follows 
$$\mathbb{E}\Big(|\mathbb{M}(n)|^{2} ; m,n,\vec{\pi},k\Big) >\frac{\Gamma(m)2^{k}\Gamma(n + 2 ||\vec{\pi}||)}{\Gamma(n)\Gamma(m + 2 ||\vec{\pi}||)} -  \frac{\Gamma(m)}{\Gamma(n)} \sum_{j=1}^{k} \frac{2^{k-1} \Gamma(n + 2 ||\vec{\pi}|| - \pi_{j})}{ \Gamma(m + 2 ||\vec{\pi}|| - \pi_{j})}. $$
\end{proof}

Bounds are now obtained for the second moment when $n$ is large in order to display explicitly the growth rate with respect to $n$. If $\vec{\pi} = \vec{0}$ the inheritance process is trivial since no there is no chance that an inheritance event can occur and $|\mathbb{M}(n)| = 1$. If $\pi_{i} = 0$ for some $i$, $1 \leq i \leq k$, then the only member of the family that occurs in instances of $\mathbb{M}$ is the original gene, and the $i^{th}$ family is non-reproductive. 

Now consider a modified version of the inheritance and duplication process. Suppose $k^{\prime}$ is the size of a subnetwork motif and $\vec{\pi}^{\prime}$ is the vector of inheritance probabilities. Then $\mathbb{M}^{\prime}$ is a subnetwork motif of size $k^{\prime}$ where all non-reproductive families have been removed from $\vec{\pi}$. If $\mathscr{I}$ is a potential instance of $\mathbb{M}$ then $\mathscr{I}^{\prime}$ is a potential instance of $\mathbb{M}^{\prime}$ which is obtained by dropping all non-reproductive genes. Then the probability of the modified inheritance event $\mathscr{I}^{\prime} \in \mathbb{M}^{\prime}(n)$ is the same as the probability of the original inheritance event $\mathscr{I} \in \mathbb{M}(n)$. Thus it suffices to study $\mathbb{M}^{\prime}$ since the probability distributions of $|\mathbb{M}(n)|$ and $|\mathbb{M}^{\prime}(n)|$ are the same. Therefore without loss of generality we can assume that $\vec{0} < \vec{\pi} \leq \vec{1}$. That is, $\mathbb{M}^{\prime}$ has no family with a zero probability of inheritance. 

\begin{corollary}
Suppose $\mathbb{M}$ is a subnetwork motif of size $k$, $1 \leq k \leq m \leq n, n \geq k^{2}$ and $\vec{0} < \vec{\pi} \leq \vec{1}$. Then the second moment of the number of instances of subnetwork motifs of size $k$ in partial binary duplication satisfies 

$$\mathbb{E}\Big(|\mathbb{M}(n)|^{2} ; m,n,\vec{\pi},k\Big) \geq  \frac{n^{2||\vec{\pi}||}2^{k}\Gamma(m)}{\Gamma(m + 2 ||\vec{\pi}||)} - \frac{1}{\Gamma(m)}\sum_{j=1}^{k} \frac{2^{k-1} n^{2 ||\vec{\pi}|| - \pi_{j}}}{ \Gamma(m + 2 ||\vec{\pi}|| - \pi_{j})}$$
and 
$$\mathbb{E}\Big(|\mathbb{M}(n)|^{2} ; m,n,\vec{\pi},k\Big) \leq \frac{n^{2||\vec{\pi}||}2^{k}\Gamma(m)}{\Gamma(m + 2 ||\vec{\pi}||)}\Big(1+\frac{6||\vec{\pi}||^{2}}{4n} \Big).$$

\end{corollary}

\begin{proof}
Since 

$$\frac{\Gamma(n+ 2 ||\vec{\pi}||)}{\Gamma(n)} = (n + 2 ||\vec{\pi}|| - 1)_{2 ||\vec{\pi}||}$$

\noindent and 

$$\frac{\Gamma(n+ 2 ||\vec{\pi}|| - \pi_{j})}{\Gamma(n)} = (n + 2 ||\vec{\pi}|| - \pi_{j} - 1)_{2 ||\vec{\pi}|| - \pi_{j}}$$

\noindent the bounds for the corollary fall directly from applying Lemma \ref{lemma3} (occurs in Section \ref{chapter:formulae}) to the result of Corollary \ref{BIbounds}.
\end{proof}

\section{Full Duplication for Multiple Subnetwork Motifs}

In this section of the thesis we will study the number of instances of subnetwork motifs in Full Duplication when there are $r$ individual subnetwork motifs before the start of the duplication. We will only explore mutliple motifs in Full Duplication in this thesis. Results for multiple subnetworks in Partial Duplication can be obtained. However, this is something that will be explored in the future. 

In this scenario, the duplication process begins with $m$ genes and ends with $n$ total genes. Initially, there are $r$ subnetwork motif instances denoted $\mathbb{M}_{i}$ for $i=1,...r$ and $\mathbb{M}_{i}(n)$ is the set of subnetwork motif instances after $n-m$ duplications for $i=1,...,r$. As before, prior to any duplications each $|\mathbb{M}_{i}(m)| = 1$. We will let $k_{i}$ be the size of $\mathbb{M}_{i}$ for $i=1, ..., r$ such that $\vec{k} = (k_{1}, ..., k_{r})$ is the vector of $r$ subnetwork motif sizes.  If we let $\mathbb{X} = \sum_{1}^{r}|\mathbb{M}_{i}(n)|$ then the expected number of subnetwork motif instances can be denoted $\mathbb{E} \Big(\mathbb{X} | m, n\Big)$.

\begin{theorem}
Let $r \geq 1$ and $k_{i} \geq 1$ for $i=1,...,r$ be as above and let $m \leq n$ such that $k_{i} \leq m$ for $i=1,...r$. Then the expected total number of subnetwork motif instances in Full Duplication is

$$\mathbb{E} \Big(\mathbb{X} | m, n\Big) = \sum_{i=1}^{r} \frac{\Gamma(n+k_{i})\Gamma(m)}{\Gamma(n)\Gamma(m+k_{i})}.$$
\end{theorem}

\begin{proof}
The lemma follows from Theorem $\ref{full_dup_fixed_n_1st_mom}$ and the linearity of the expectation.
\end{proof}

Now we are going to derive the second moment of $\mathbb{X}$. In order to do so we will apply linearity of expectation to $\mathbb{X}^{2}$ which gives 

\begin{align}\label{ds}
\mathbb{E}\Big(\mathbb{X}^{2} ; m,n\Big) = \sum_{i=1}^{r} \mathbb{E}\big(|\mathbb{M}_{i}(n)| \cdot |\mathbb{M}_{i}(n)|\big) + 2\sum_{i=1}^{r-1} \sum_{j=i+1}^{r} \mathbb{E}\big(|\mathbb{M}_{i}(n)|\cdot|\mathbb{M}_{j}(n)|\big).
\end{align} 

Note that the summand in the first summation is given by Theorem $\ref{full_dup_fixed_n_2nd_mom}$. In order to illustrate what may occur in the second summation we will consider the example $r=2, k_{1} = 2$, and $k_{2} =2$. Then there are three different scenarios for the joint distribution of $\mathbb{M}_{1}(n)$ and $\mathbb{M}_{2}(n)$. Each scenario differs based on how the 2 original subnetwork motifs are related. The two individual subnetwork motif instances can present as disjoint such that $\mathscr{I}_{1} = (a_{1}, a_{2}) \in \mathbb{M}_{1}(n)$ and $\mathscr{I}_{2} = (a_{3}, a_{4}) \in \mathbb{M}_{2}(n)$. In the second scenario the subnetwork motifs can share one gene such that $\mathscr{I}_{1} =  (a_{1}, a_{2}) \in \mathbb{M}_{1}(n)$ and $\mathscr{I}_{2} =  (a_{1}, a_{3}) \in \mathbb{M}_{2}(n)$. Lastly, the subnetwork motifs can share the same genes with differing mechanisms where $\mathscr{I}_{1} = (a_{1}, a_{2}) \in \mathbb{M}_{1}(n)$ and $\mathscr{I}_{2} = (a_{1}, a_{2}) \in \mathbb{M}_{2}(n)$ are original instances of these genes. In order to calculate the results for $\mathbb{E}\big(|\mathbb{M}_{1}(n)| \cdot |\mathbb{M}_{2}(n)|\big)$ we will use $a(x), b(x), \textrm{ and } c(x)$ from the proofs of Theorems 1 and 2. The results for each case are as follows:

Shared gene case where $m=3$:
\begin{align*}
\mathbb{E}\big(|\mathbb{M}_{1}(n)|\cdot|\mathbb{M}_{2}(n)|\big) &= [x^{n}]\Big(a(x)^{m-3} \cdot b(x)^{2} \cdot c(x)\Big)\binom{n-1}{m-1}^{-1}\\ 
&= \Big([x^{n-m-1}]2(1-x)^{-(m+4)} + [x^{n-m}](1-x)^{-(m+3)}\Big)\binom{n-1}{m-1}^{-1} \\ 
&= \Big(\frac{2\Gamma(n+3)\Gamma(m)\Gamma(n-m+1)}{\Gamma(m+5)\Gamma(n-m-1)\Gamma(n)} + \frac{\Gamma(n+3)\Gamma(m)\Gamma(n-m+1)}{\Gamma(m+4)\Gamma(n-m)\Gamma(n)}\Big).
\end{align*}

Same genes, different mechanism case where $m=2$:
\begin{align*}
\mathbb{E}\big(|\mathbb{M}_{1}(n)|\cdot|\mathbb{M}_{2}(n)|\big) &= [x^{n-m-i}]\sum_{i=0}^{2} 2^{i}  \binom{2}{i} (1-x)^{-(m+i+2)}\binom{n-1}{m-1}^{-1} \\ 
&= \sum_{i=0}^{2} 2^{i}\binom{2}{i} \frac{\Gamma(n+2)\Gamma(n-m+1)\Gamma(m)}{\Gamma(n-m-i+1)\Gamma(m+i+2)\Gamma(n)}.
\end{align*}

We will now look at the expectation of the number of instances of $r$ subnetwork motifs. Initially there are $r$ original subnetwork motif instances $\mathscr{I}_{1}, ..., \mathscr{I}_{r}$ with size $k_{1},...,k_{r}$. Suppose for original instances $\mathscr{I}_{i}$ and $\mathscr{I}_{j}$ there exist an overlap in the genes that make up each instance, then we denote the overlap size $\ell_{i,j}$. Then the summand in the first summation of Equation \ref{ds} is a function of $m,n,k_{i},k_{j}\textrm{ and }\ell_{i,j}$. Thus 

$$\mathbb{E}\Big(|\mathbb{M}_{i}(n)|\cdot|\mathbb{M}_{j}(n)|\Big) = \binom{n-1}{m-1}^{-1}[x^{n}]\Big(a(x)^{m-k_{i}-k_{j}}b(x)^{k_{i}+k_{j}-\ell_{i,j}}(c(x)^{\ell_{i,j}}\Big).$$

\begin{lemma}\label{exp_of_two_motifs}
Suppose $\mathbb{M}_{i}$ and $\mathbb{M}_{j}$ are subnetwork motifs of size $k_{i}$ and $k_{j}$ respectively, $l_{i,j}$ is the overlap in genes between $\mathbb{M}_{i}$ and $\mathbb{M}_{j}$, and $1\leq k_{i} \leq k_{j} \leq m \leq n$. Then $\mathbb{E}\Big(|\mathbb{M}_{i}(n)|\cdot|\mathbb{M}_{j}(n)|\Big)$ is the expected number of instances of $\mathbb{M}_{i}$ and $\mathbb{M}_{j}$, which evaluates to
$$\frac{\Gamma(n+k_{i}+k_{j})\Gamma(m)\Gamma(n-m+1)}{\Gamma(n)}\mathlarger{\mathlarger{\sum}}_{u=0}^{\ell_{i,j}}2^{u}\binom{\ell_{i,j}}{u}\Big(\Gamma(m+k_{i}+k_{j}+u)\Gamma(n-m-u+1)\Big)^{-1}.$$
\end{lemma}

\begin{proof}

$$[x^{n}]\Big(\frac{x}{1-x}\Big)^{m-k_{i}-k_{j}}\Big(\frac{x}{(1-x)^{2}}\Big)^{k_{i}+k_{j}-\ell_{i,j}}\Big(\sum_{i=1}^{\ell_{i,j}} \binom{\ell_{i,j}}{i}(\frac{2x^{2}}{(1-x)^{3})^{i}}(\frac{x}{(1-x)^{2}})^{\ell_{i,j}}\Big)$$

\end{proof}

In the following theorem we define $R(m,n)$ as $\frac{\Gamma(m)\Gamma(n-m+1)}{\Gamma(n)}$.

\begin{corollary}\label{ordered_pairs_multi_full}
Let $r \geq 1$ and $k_{i} \geq 1$ be the size of $\mathbb{M}_{i}$ for $i=1, ..., r$. Then the expected number of ordered pairs of subnetwork motifs in Full Duplication is $\mathbb{E} \bigl(\mathbb{X}^{2} \,|\, m, n\bigr)$, which evaluates to $R(m,n)$ times 
\begin{multline*}
\sum_{i=1}^{r} \Bigl(\Gamma(n+k_{i}) \mathlarger{\sum}_{u=0}^{k_{i}}2^{u} \binom{2}{u} \bigl(\Gamma(m+{2+u})\Gamma(n-m-u+1)\bigr)^{-1} \\
+ 2\sum_{i=1}^{r-1}\sum_{j=i+1}^{r} \Gamma(n+k_{i}+k_{j})\mathlarger{\sum}_{u=1}^{\ell_{i,j}}2^{u}\binom{\ell_{i,j}}{u}\bigl(\Gamma(m+k_{i}+k_{j}+u)\Gamma(n-m-u+1)\bigr)^{-1}\Bigr).
\end{multline*}
\end{corollary}

\begin{proof}
The corollary follows from applying Theorem \ref{full_dup_fixed_n_2nd_mom} and Lemma \ref{exp_of_two_motifs} to Equation \ref{ds}. 
\end{proof}

\section{Useful Formulae for Approximating Moments}
\label{chapter:formulae}

We will now derive approximations that are both rapid in computation and maintain accuracy for large $n.$ Note that large $n$ corresponds in biology to an organism with a large genome. The following lemmas were useful for exponentiating approximations to logarithms for various moments in the previous section.

\begin{lemma}\label{explemma}
If $0 \leq \zeta \leq \frac{1}{2}$ then $$1 + \zeta \leq e^{\zeta} \leq 1 + \frac{3\zeta}{2}.$$
\end{lemma}

\begin{proof}
We know that for any $\zeta$ 
$$e^{\zeta} = \sum_{i=0}^{\infty} \frac{\zeta^{i}}{i!}.$$ The lower bound is obvious for any $\zeta \geq 0$. For the upper bound we have $$e^{\zeta} \leq 1 + \zeta + \zeta \sum_{i=2}^{\infty} \frac{(1/2)^{i-1}}{2} = 1 + \frac{3}{2}\zeta.$$
\end{proof}

\begin{lemma}\label{negexplemma}
If $0 \leq \zeta \leq \frac{1}{2}$ then $$1 - \zeta \leq e^{-\zeta} \leq 1 - \zeta + \frac{1}{2}\zeta^{2}.$$
\end{lemma}

\begin{proof}
We know that 
$$e^{-\zeta} = \sum_{i=0}^{\infty} \frac{(-1)^{i}\zeta^{i}}{i!}.$$ Since $|\zeta| \leq \frac{1}{2}$, the terms in the series expansion of $e^{-\zeta}$ are alternating in sign and their absolute value is strictly decreasing and tending to 0. The lower bound is obtained by truncating the series sum after a negative term and similarly the upper bound is obtained by truncating the series sum after a positive term. 
\end{proof}


\begin{lemma}\label{lemma3}
If $j \geq 1$ and $y \geq j^{2}$ then 
$$1 \leq \frac{(y+j-1)_{j}}{y^{j}} \leq 1 + \frac{3j^{2}}{4y}.$$
\end{lemma}

\begin{proof}
Since the lower bound is trivial we will focus on bounding the ratio above. Let 
$$\delta = \ln(\frac{(y+j-1)_{j}}{y^{j}}) = \ln\Big( \prod_{r=1}^{j-1}(1+\frac{r}{y}) \Big).$$ Since the log of the product is equal to the sum of the logs and $\ln(1+\frac{r}{y}) \leq \frac{r}{y}$ for $r \geq 0$ we obtain the following bounds \begin{equation}\label{lemma3eq}\delta \leq \sum_{r=1}^{j-1} \Big(\frac{r}{y}\Big) = \frac{\binom{j}{2}}{y} \leq \frac{j^{2}}{2y} \leq \frac{1}{2}.\end{equation}

Note that $e^{\delta} \leq 1 + \frac{3}{2}\delta$ whenever $0 \leq \delta \leq \frac{1}{2}$ and  $\frac{j^{2}}{2y} \leq \frac{1}{2}$, therefore 
$$e^{\delta} = \sum_{i=0}^{\infty} \frac{\delta^{i}}{i!} \leq 1 + \delta + \delta \sum_{i=2}^{\infty} \frac{(1/2)^{i}}{i!} \leq 1 + \frac{3}{2}\delta$$

Since $\frac{j^{2}}{2y} \leq \frac{1}{2}$ we can apply Lemma \ref{explemma} to obtain the result of the lemma. 
\end{proof}

\begin{lemma}\label{lemma4}
If $j \geq 1$ and $y \geq j^{2}$ then

$$1 - \frac{j^{2}}{2y} \leq \frac{y^{j}}{(y+j-1)_{j}} \leq 1.$$
\end{lemma}

\begin{proof}
Since the upper bound is trivial we will focus on bounding the ratio below. Using $\delta$ as it is in Lemma,
note that $e^{-\delta} \geq 1 - \delta$ for any real $\delta$.

Now we know $\delta \leq \frac{j^{2}}{2y}$. Thus we will bound $e^{-\delta}$ to obtain the following
$$e^{-\delta} \geq 1 - \frac{j^{2}}{2y} \textrm{ when $\delta \geq 0$}.$$
\end{proof}

\begin{lemma}\label{lemma5} Suppose $j \geq 1$ and $y \geq 2j$ then $$1 - \frac{j^{2}}{y} \leq \frac{(y)_{j}}{y^{j}} \leq 1.$$ \end{lemma}

\begin{proof}
Since the upper bound is trivial we will focus on bounding the expression below. Let
$$\delta = - \ln(\frac{(y)_{j}}{y^{j}}) = - \ln(\prod_{r=0}^{j+1}(1 - \frac{r}{y}))$$

In order to prove the statement of the Lemma we must obtain bounds for the above expression. Since $|\frac{r}{y}| < 1$ we will use the Taylor Series to obtain bounds. Let 

$$\delta =  \sum_{r = 1}^ { j - 1} \sum_{i=1}^{\infty} \frac{(r/y)^{i}}{i} =   \sum_{i=1}^{\infty}\sum_{r = 1}^ { j - 1} \frac{(r/y)^{i}}{i}.$$ 

When $i=1$ we know that 

$$\sum_{r =1}^{j-1} \frac{r}{y} = \frac{1}{y}(1 + 2 + ... + j-1) = \frac{j(j-1)}{2y} \leq \frac{j^2}{2y}.$$ 

When $i \geq 2$ a bound for the inner summation is as follows $$\sum_{r = 1}^ { j - 1} \frac{(r/y)^{i}}{i} \leq \frac{j^{i+1}}{2y^{i}}.$$ 

Then

 $$\sum_{i=2}^{\infty}  \frac{j^{i+1}}{2y^{i}} = \frac{j^{2}}{2y}  \sum_{i=2}^{\infty} (\frac{j}{y})^{i-1} \leq \frac{j^{2}}{2y} \sum_{i=2}^{\infty} (\frac{1}{2})^{i-1} = \frac{j^{2}}{2y}$$ 
 
since
 
$$\sum_{i=1}^{\infty} (\frac{1}{2})^{i-1} = 1.$$ 

Thus the lemma follows since 

$$0 \leq \delta \leq \frac{j^{2}}{y}$$ 

so that 

$$1 - \frac{j^{2}}{y} \leq 1 - \delta \leq e^{-\delta} \leq 1.$$

\end{proof}

\begin{lemma} \label{lemma6}
Suppose $j \geq 1$ and $y \geq 2j^{2}$ then 
$$1 \leq \frac{y^{j}}{(y)_{j}} \leq 1 + \frac{3j^{2}}{2y}.$$
\end{lemma}

\begin{proof}
Since the lower bound is trivial we will focus on bounding the expression above. Let
$$\frac{y^{j}}{(y)_{j}} = e^{\delta}$$ 
where $\delta$ is as is in the proof of Lemma \ref{lemma5}.

From the proof of Lemma \ref{lemma5} we know that 

$$\delta \leq \frac{j^{2}}{y}.$$

Therefore $\delta \leq \frac{1}{2}$ since $y \geq 2j^{2}$.

From proof of Lemma \ref{lemma3} we know that

$$e^{\delta} \leq 1 + \frac{3}{2}\delta \textrm{ when } \delta \leq \frac{1}{2}$$

Thus 

$$1 \leq e^{\delta} \leq 1 + \frac{3}{2}\delta \leq 1 + \frac{3j^{2}}{2y}.$$
\end{proof}

\begin{lemma}\label{lemma10}
If $j \geq 1$ and $y \geq 2j^{2}$ then 
$$1 \leq \Big( \frac{(y+j-1)_{j}}{y^{j}}\Big)^{2} \leq 1 + \frac{3j^{2}}{2y}.$$
\end{lemma}

\begin{proof}
In order to prove the statement of the lemma let 
$$\delta = \ln\Big((\frac{(y+j-1)_{j}}{y^{j}})^{2}\Big) = 2\ln\Big( \prod_{r=1}^{j-1}(1+\frac{r}{y}) \Big),$$

just as in the proof of Lemma 3. Therefore the ratio we are bounding is $e^{2\delta}$. Recall from Equation \ref{lemma3eq} that $\delta \leq \frac{j^{2}}{2y}$ and $2\delta \leq 1/2$ which mean $\delta \leq 1/4$.
Thus, 

$$e^{2\delta} \leq 1 + 3\delta \leq 1 + \frac{3j^{2}}{2y}$$ since $e^{\delta} \leq 1 + \frac{3}{2} \delta.$
\end{proof}

\section{Discussion}

\subsection{Connection to Polya Urns}

To study a more realistic model it is best to generalize the previous duplication model by generalizing the start of the duplication process. The $\textit{gene}$ $\textit{duplication}$ $\textit{process}$ discussed in Section \ref{chapter:motifs} is an extension of the random duplication process that begins with $w$ individual gene families where each family has size $s_{i} \geq 1$ for $i=1, ..., w.$ We let $\vec{s} = (s_{1}, ..., s_{w})$ and $m = s_{1} + ... + s_{w}$. At each step, a random gene is selected to be duplicated. If a gene in the $i^{th}$ family is duplicated then the new duplicated gene belongs to the $i^{th}$ family. After $n-m$ duplications we let $t_{i} \geq 0$ be the number of duplicated genes that have been added to the $i^{th}$ family and $\vec{t} = (t_{1}, ..., t_{w})$. Note that the gene duplication process is a special case of this generalized process that begins with $m$ single gene families.

\begin{theorem}\label{kriz}
Let $\vec{s}$ be the initial composition of family sizes and let $\vec{X}$ be the vector of numbers of individuals added to each family after $n - m \geq 0$ random duplications. Then for any weak composition $\vec{t}$ of $n-m$ into $w$ parts we have $$P[\; \vec{X} = \vec{t} \; | \; \vec{s}\;] = \binom{n- m}{t_{1}, ..., t_{w}} \frac{(m-1)!}{(n-1)!} {\prod_{j=1}^{w} \frac{(s_{j} + t_{j} -1)!}{(s_{j} -1)!}}.$$
\end{theorem}

\begin{proof}
We need to calculate the likelihood that $\vec{X} = \vec{t}$ given $\vec{s}$. In order for $\vec{X} = \vec{t}$ at the end of a duplication sequence, the number of genes that have been duplicated in the $j^{th}$ family must be $x_{j}$ for $1 \leq j \leq w$. That is to say that $x_{j} = t_{j}$ for $j=1, ..., w$ after $n-m$ duplications.  There are exactly $t_{j}$ locations in the sequence of $n-m$ duplications in which the duplication can occur in the $j^{th}$ family for $1\leq j \leq w$. Then a standard combinatorial fact is that the number of ways we can choose the locations so that $\vec{X} = \vec{t}$ is the multinomial coefficient $\binom{n-m}{t_{1}, ..., t_{w}}.$

Given one of these patterns we can calculate the probability of its occurrence in a random duplication process as follows. Consider the $i^{th}$ duplication in the series of $n-m$ duplications. For $\vec{X} = \vec{t}$, the given pattern determines the $j$ such that the $j^{th}$ family that must be duplicated at the $i^{th}$ step. Then the probability that the $i^{th}$ duplication occurs in the $j^{th}$ family is the current size of the $j^{th}$ family divided by $n+i-1$, since the total number of genes at the $i^{th}$ duplication step is $n+i-1$. Therefore, the probability that  $\vec{X} = \vec{t}$ is the product of the probabilities that the $i^{th}$ duplication occurs in the $j^{th}$ family for $i=1, ..., n-m$ and $j=1, ..., w$. Then for $i=1, ..., n-m$, the denominators are as follows

$$\Big((m)\cdot (m+1) \cdot \cdots \cdot (m+(n-m-1))\Big) = \frac{(n-1)!}{(m-1)!}$$

\noindent and for $j=1,...,w$ the numerators are

$$\Big({s_{j}} \cdot ({s_{j} + 1}) \cdots  ({s_{j}+t_{j}-1})\Big) = \frac{(s_{j} + t_{j} -1)!}{(s_{j} -1)!}.$$ 

Therefore the probability that the given pattern is actually followed such that $\vec{X} = \vec{t}$ is

$$\frac{(m-1)!}{(n-1)!} {\prod_{j=1}^{w} \frac{(s_{j} + t_{j} -1)!}{(s_{j} -1)!}}.$$

Clearly the $\binom{n-m}{t_{1}, ..., t_{w}}$ different patterns of duplication are mutually exclusive. Therefore the total probability of $\vec{X} = \vec{t}$ given $\vec{s}$ is the product of this multinomial coefficient and the probability of duplication for each family, giving the result of the theorem. 
\end{proof}

The formula from the conclusion of Theorem \ref{kriz} shows up in the analysis of the Kriz polya urn model \citep{polyaurn} in the case that the Kriz model parameter $s=1$. The Kriz model is a multi-urn model used to study the spread of a disease and the parameter $s$ gives the number of people likely to to come into contact with the disease in a specific unit of time. The result of Theorem $\ref{kriz}$ can be expressed as 
$$P[\; \vec{X} = \vec{t} \; | \; \vec{s}\;] = \binom{n- m}{t_{1}, ..., t_{w}}\frac{\prod_{j=1}^{w}(\prod_{l=0}^{t_{j}-1}(s_{j} + l))}{\prod_{i=m}^{n-1}i},$$
\noindent which is essentially the same formula as the one following the displayed expression \citep[eq. (9.8)]{polyaurn} after setting the Kriz model parameter $s$ to 1.

The Kriz model is expressed in terms of urns, balls, and colors. To correspond to our model, the urn would be the entire genome, the balls would be the genes, and the colors would be the families. Our model is a special case of the Kriz model where there is one urn with $n$ balls, $m$ colors, and only one ball can be added at each step. 

\subsection{Connection to MITES}

It has been expressed that transposable elements are one of the reasons a genome can make programmed responses to environmental challenges \citep{mcclintock}. At its core, the genome uses transposable elements to rewire expressions in response to environmental challenges \citep{mcclintock}. The rice genome has putative regulatory elements called MITES also know as miniature inverted transposable elements \citep{wessler} that identify which subnetwork motifs are may be under selection to an environmental challenge. These MITES undergo a high rate of duplication and genes in the rice genome are sorted based on their MITE profile. This results in MITE-infested genes having a history of regulation by their MITES.

If two genes A and B share the same MITE, they are said to share a regulatory link. In the rice genome a simple subnetwork motif would be a connected collection of three genes and three MITES, where genes A and B share MITE 1, genes A and C share MITE 2, and genes B and C share MITE 3.

Genes in the rice genome can be sorted into gene families based on their DNA sequence or their MITE profile. The age of the links between genes can be characterized from the sequences of each MITE  \citep{bennetzen}. That is to say, the more divergence there is in the sequence between MITEs, the older the link between the genes. Moreover, the MITEs associated with genes can define regulatory links. Therefore, a history of different genes and their potential regulation by MITES can be reconstructed \citep{bennetzen}.

The modeling of MITE regulatory evolution is a direct application of the duplication models discussed in this thesis. Suppose we have the simple network discussed in the previous paragraph. If gene A gets duplicated to create gene A’ and the MITE between A’ and C is retained but the MITE between A’ and B is not, then the simple regulatory network has undergone Partial Duplication.

Subnetwork motifs can be examined in the rice genome using MITES since it should be possible to consider a couple of gene families that have expanded over time and how the MITES evolved with them. To a limited extent the fate of the network can be examined. 

\subsection{Mixed Duplication Model}

The mixed duplication model defined in \citep{fanchung} is a duplication model where nodes in a network are duplicated along with their connections, which is a common occurrence in biological systems, especially in the context of gene duplication \citep{fanchung}. In the model the results of both Full Duplication and Partial Duplication on the structure and connectivity of the network are examined. Note that node connectivity refers to the regulatory relation between genes in a genome. Under the mixed duplication model both full and Partial Duplication are considered at the same time. There are two probabilities, $p$ and $q$, that define the behavior of the model. At each step in the duplication process a node is randomly selected for duplication. Then with probability $1-q$ all regulatory links are retained and with probability $q$ each link is retained independently with probability $p$.

What we consider Full Duplication can be described as mixed duplication when $q=0$, which is how it is referenced in \citep{fanchung}. However, our Partial Duplication mode is a vast generalization of the mixed duplication model in \citep{fanchung} such that each family has its own inheritance probability for regulatory links. One of the limitations of the mixed duplication model is that it is restrictive to situations where all regulation relationships have the same probability of inheritance through duplication. One case in which this model can be applied is when transcription factors are regulating a very high number of targets and have independently mutating binding sites. Our model extends the applications of the mixed duplication model and is also appropriate for transcription factors that are regulating several targets and the mutations occur in the regulator rather than the binding site. 

The mixed duplication model is a special case of our Binary Inheritance mode where the inheritance probability for each $\pi_{i}$ for $i=1, ..., k$ is $\pi_{i} = 1 - q + qp$. Consider the mixed duplication model where the probability for link retention under Partial Duplication is $p=0$. Then at each step of the node duplication process $1-q$ is the probability that all of the links where retained in Full Duplication and $q$ is the probability that none of the links where retained in Partial Duplication. As already noted, the subnetwork motif results of this thesis apply to the mixed duplication model \citep{fanchung}.

\bibliographystyle{splncs041}
\bibliography{thesis.bib}
 \addcontentsline{toc}{chapter}{Bibliography}

\end{document}